\documentclass[longauth]{aa}

\usepackage{graphicx}
\usepackage{url}
\usepackage{multirow}
\usepackage{natbib}
\bibpunct{(}{)}{;}{a}{}{,}
\usepackage{float}

\DeclareFontFamily{U}{euc}{}
\DeclareFontShape{U}{euc}{m}{n}{<-6>eurm5<6-8>eurm7<8->eurm10}{}%
\DeclareSymbolFont{AMSc}{U}{euc}{m}{n} 
\DeclareMathSymbol{\umu}{\mathord}{AMSc}{"16}

\newcommand{\specialcell}[2][c]{%
  \begin{tabular}[#1]{@{}c@{}}#2\end{tabular}}

\begin{document}
\title{Azimuthal asymmetries in the debris disk around HD\,61005\thanks{Based on observations made with ESO Telescopes at the Paranal Observatory under programs ID 095.C-0298 and 095.C-0273. Data of Fig.\,1 is only available in electronic form at the CDS via anonymous ftp to cdsarc.u-strasbg.fr (130.79.128.5) or via http://cdsweb.u-strasbg.fr/cgi-bin/qcat?J/A+A/. Based on \textit{Herschel} observations, OBSIDs: 1342270977, 1342270978, 1342270979, 1342270989, and 1342255147. \textit{Herschel} is an ESA space observatory with science instruments provided by European-led Principal Investigator consortia and with important participation from NASA.}}
\subtitle{A massive collision of planetesimals ?}

\author{J. Olofsson\inst{1, 2, 3}
    \and
    M. Samland\inst{3}
    \and
    H. Avenhaus\inst{4, 2}
    \and
    C. Caceres\inst{1,2}
    \and
    Th. Henning\inst{3}
    \and
    A. Mo\'or\inst{5}
    \and
    J. Milli\inst{6}
    \and
    H. Canovas\inst{1,2}
    \and
    S. P. Quanz\inst{7}
    \and
    M. R. Schreiber\inst{1,2}
    \and
    J.-C. Augereau\inst{8, 9}
    \and
    A. Bayo\inst{1, 2}
    \and
    A. Bazzon\inst{7}
    \and
    J.-L. Beuzit\inst{8, 9}
    \and
    A. Boccaletti\inst{10}
    \and
    E. Buenzli\inst{7}
    \and
    S. Casassus\inst{4, 2}
    \and
    G. Chauvin\inst{8, 9}
    \and
    C. Dominik\inst{11}
    \and
    S. Desidera\inst{12}
    \and
    M. Feldt\inst{3}
    \and
    R. Gratton\inst{12}
    \and
    M. Janson\inst{3, 13}
    \and
    A.-M. Lagrange\inst{8, 9}
    \and
    M. Langlois\inst{14, 15}
    \and
    J. Lannier\inst{8, 9}
    \and
    A.-L. Maire\inst{12, 3}
    \and
    D. Mesa\inst{12}
    \and
    C. Pinte\inst{16, 8}
    \and
    D. Rouan\inst{10}
    \and
    G. Salter\inst{15}
    \and
    C. Thalmann\inst{7}
    \and
    A. Vigan\inst{15, 6}
}

\institute{
Instituto de F\'isica y Astronom\'ia, Facultad de Ciencias, Universidad de Valpara\'iso, Av. Gran Breta\~na 1111, Playa Ancha, Valpara\'iso, Chile\\\email{johan.olofsson@uv.cl}
\and
ICM nucleus on protoplanetary disks, ``Protoplanetary discs in ALMA Early Science'', Chile
\and
Max Planck Institut f\"ur Astronomie, K\"onigstuhl 17, 69117 Heidelberg, Germany
\and
Departamento de Astronom\'ia, Universidad de Chile, Casilla 36-D, Santiago, Chile
\and
Konkoly Observatory, Research Centre for Astronomy and Earth Sciences, Hungarian Academy of Sciences, PO Box 67, H-1525 Budapest, Hungary
\and
European Southern Observatory (ESO), Alonso de C\'ordova 3107, Vitacura, Casilla 19001, Santiago, Chile
\and
Institute for Astronomy, ETH Zurich, Wolfgang-Pauli-Strasse 27, CH-8093 Zurich, Switzerland
\and
Univ. Grenoble Alpes, Institut de Plan\'etologie et d'Astrophysique de Grenoble (IPAG, UMR 5274), F-38000 Grenoble, France
\and
CNRS, Institut de Plan\'etologie et d'Astrophysique de Grenoble (IPAG, UMR 5274), F-38000 Grenoble, France
\and
LESIA, Observatoire de Paris, CNRS, Universit\'e Pierre et Marie Curie 6 and Universit\'e Denis Diderot Paris 7, 5 place Jules Janssen, 92195 Meudon, France
\and
Astronomical Institute Anton Pannekoek, University of Amsterdam, PO Box 94249, 1090 GE Amsterdam, The Netherlands
\and
INAF Osservatorio Astronomico di Padova, Vicolo dell'Osservatorio 5, 35122 Padova, Italy
\and
Department of Astronomy, Stockholm University, AlbaNova University Center, 10691 Stockholm, Sweden
\and
CNRS, Centre de Recherche Astrophysique de Lyon, Observatoire de Lyon, Ecole Normale Sup\'erieure de Lyon, Universit\'e Lyon 1, 9 avenue Charles Andr\'e, Saint-Genis Laval, 69230, France
\and
Aix Marseille Universit\'e, CNRS, LAM (Laboratoire d'Astrophysique de Marseille) UMR 7326, 13388, Marseille, France
\and
UMI-FCA, CNRS/INSU France, and Departamento de Astronom\'ia, Universidad de Chile, Casilla 36-D Santiago, Chile
}

\abstract{Debris disks offer valuable insights into the latest stages of circumstellar disk evolution, and can possibly help us to trace the outcomes of planetary formation processes. In the age range 10 to 100\,Myr, most of the gas is expected to have been removed from the system, giant planets (if any) must have already been formed, and the formation of terrestrial planets may be on-going. Pluto-sized planetesimals, and their debris released in a collisional cascade, are under their mutual gravitational influence, which may result into non-axisymmetric structures in the debris disk.}
{High angular resolution observations are required to investigate these effects and constrain the dynamical evolution of debris disks. Furthermore, multi-wavelength observations can provide information about the dust dynamics by probing different grain sizes.}
{Here we present new VLT/SPHERE and ALMA observations of the debris disk around the 40\,Myr-old solar-type star HD\,61005. We resolve the disk at unprecedented resolution both in the near-infrared (in scattered and polarized light) and at millimeter wavelengths. We perform a detailed modeling of these observations, including the spectral energy distribution.}
{Thanks to the new observations, we propose a solution for both the radial and azimuthal distribution of the dust grains in the debris disk. We find that the disk has a moderate eccentricity ($e \sim 0.1$) and that the dust density is two times larger at the pericenter compared to the apocenter.}
{With no giant planets detected in our observations, we investigate alternative explanations besides planet-disk interactions to interpret the inferred disk morphology. We postulate that the morphology of the disk could be the consequence of a massive collision between $\sim$\,1000\,km-sized bodies at $\sim$\,61\,au. If this interpretation holds, it would put stringent constraints on the formation of massive planetesimals at large distances from the star.}


\titlerunning{The asymmetric disk around HD\,61005}

\maketitle

\section{Introduction}\label{sec:introduction}

Debris disks are the leftovers of star and planetary formation processes (see \citealp{Wyatt2008,Krivov2010,Matthews2014} for recent reviews). Departure from photospheric emission at infrared (IR) wavelengths was first discovered around Vega (\citealp{Aumann1984}) using the Infrared Astronomical Satellite (IRAS). This excess emission was originally thought to be the remnant of the cloud out of which Vega formed. Several decades later, we now know that the dusty ``debris'' responsible for the IR emission are arranged in a disk comparable to the Edgeworth-Kuiper belt in the solar system. The dust grains, with sizes between a few $\umu$m to a few millimeters, located at tens of au from the central star are heated by the stellar radiation and re-emit at mid-IR and mm wavelengths. Since the original discovery, and mostly thanks to space-based missions such as the Infrared Space Observatory, {\it Spitzer}, and {\it Herschel}, several hundred of main sequence stars are known to harbor debris disks (e.g., \citealp{Eiroa2013,Chen2014}).

Recent decades have seen incredible progress in the field of disk observations (e.g., \citealp{Augereau1999,Kalas2005,Buenzli2010,Lebreton2012,MillarBlanchaer2015}). Observations with ever improving spatial resolution have revealed asymmetric disks (e.g., \citealp{Lagrange2015}; \citealp{Kalas2015}) as well as complex, moving, small-scale structures (\citealp{Boccaletti2015}). Nonetheless, spatially resolved observations of debris disks remain relatively rare and even though there are theoretical works focusing on the dynamical evolution of debris disks (e.g., \citealp{Dominik2003}, \citealp{Kenyon2006}), they still need to be confronted with the observations. The current paradigm is that the primordial gas-rich proto-planetary disks are thought to have a half-life time of about $2-3$\,Myr (\citealp{Hernandez2007}). As a disk evolves Pluto-sized planetesimals can form (\citealp{Johansen2015}) along with giant planets which may accrete their mass from the gas reservoir (core accretion or gravitational instability scenarios). After a few Myr, the gaseous content is quickly removed from the disk by efficient processes such as photo-evaporation (e.g., \citealp{Alexander2006}; \citealp{Owen2011}). Only already formed planets, planetesimals and dust grains will thus remain while the disk enters its debris disk phase. After a short phase of runaway growth, terrestrial planets may form in the inner regions of the disk, with Pluto-sized bodies in the outer regions, via chaotic growth of these oligarchs, on a timescale of $10-100$\,Myr (\citealp{Kenyon2006,Kenyon2008,Kenyon2010}). The time evolution of the entire system then becomes more regular and less chaotic. The km-sized bodies, arranged in one or more planetesimal belt(s), evolve under their mutual gravitational influence. Through collisions, they continuously release small particles in a collisional cascade. Small dust grains, in turn, are removed from the system either by radiation pressure or Poynting-Robertson drag. Therefore, one can consider that after $\sim$\,100\,Myr, planetary formation has stopped. The system is left with one (or more) planetesimal belt(s) and, quite possibly, with planets of various masses. By observing systems in the range $10-100$\,Myr, one can therefore study the time evolution of debris disks. Spatially resolved images can provide constraints on the radial and azimuthal distribution of the dust, giving us insight about the dynamics at stake in these systems.

Here, we present {\it Very Large Telescope} (VLT) \textit{Spectro-Polarimetric High-contrast Exoplanet REsearch} (SPHERE) and Atacama Large Millimeter/submillimeter Array (ALMA) observations of the debris disk around the solar type star \object{HD\,61005} (G8V), located at a distance of $35.4\pm1.1$\,pc (\citealp{2007}). The age of the system is believed to be within $40_{-30}^{+10}$\,Myr old, based on membership of the Argus association (\citealp{Desidera2011,DeSilva2013,Elliott2014}). The uncertainties for the age are mostly related to the dispersion in ages reported in the literature for both the Argus association and the IC\,321 super cluster. In the last ten years, it has been spatially resolved on multiple occasions with several instruments; \citet[][Hubble Space Telescope HST/NICMOS]{Hines2007}; \citet[][HST/ACS]{Maness2009}; \citet[][VLT/NaCo]{Buenzli2010}; \citet[][Submillimeter Array SMA]{Ricarte2013}; and \citet[][HST/STIS]{Schneider2014}. The disk earned its nickname of The Moth because of the swept-back wings first revealed in the HST observations of \citet{Hines2007}. The wings may originate from the interaction between the interstellar medium (ISM) and the disk itself; as the star moves through the local ISM, (small) dust grains are set on eccentric orbits, drifting away from the central star.

While \citet{Hines2007} and \citet{Maness2009} mostly studied the wings, the study presented in \citet{Buenzli2010} resolved the debris disk as a ring, thanks to the better angular resolution provided by the NaCo instrument. They found the disk to be almost edge-on ($i = 84.3\pm1^{\circ}$), to have a semi-major axis of $61.25\pm0.85$\,au with an eccentricity of $e = 0.045\pm0.015$, which translates into an offset of $2.75\pm0.85$\,au of the star with respect to the disk. A brightness asymmetry between the two ansae is observed, which cannot be fully explained by the off-centering. No planets were detected by \citet{Buenzli2010}, for observations that should, in principle, have detected planets with masses starting from a few Jupiter masses. The integrated luminosity of the disk is large at IR wavelengths ($L_{\mathrm{disk}}/L_{\star} \sim 3 \times 10^{-3}$) and is best modeled by two spatially separated dust belts. The main dust belt is the one located at $\sim$60\,au presented in \citet{Buenzli2010}, but a warm component is usually required to reproduce the spectral energy distribution (SED). \citet{Ricarte2013} argue that this additional belt is mandatory to match the SED at about 20\,$\umu$m, however its location remains unconstrained. In this paper, we focus on constraining the properties of the debris disk which is resolved at unprecedented angular resolution at both near-IR and mm wavelengths. Studying the properties and origin of the swept-back wings is beyond the scope of this paper, since they are marginally detected with these observations.

\section{Observations, data processing, and stellar parameters}\label{sec:observations}

Table\,\ref{tab:log} summarizes the VLT/SPHERE and ALMA observations presented in this study.

\begin{table*}
\centering
\caption{Log for the VLT/SPHERE and ALMA observations\label{tab:log}}
\begin{tabular}{@{}lclcccc@{}}
\hline\hline
\multicolumn{7}{c}{VLT/SPHERE} \\
\hline
Observing date   & Prog. ID & Instrument Mode & Filter & Seeing & Airmass & Coherence time \\
$[$YYYY-MM-DD$]$ &          &                 &        & $[^{\prime\prime}]$ &   & $[$ms$]$ \\
\hline
2015-02-03 & 95.C-0298 & IRDIFS      & $H_2H_3$/$YJ$ & 0.67 & 1.01 & 22.0 \\
2015-03-30 & 95.C-0298 & IRDIFS\_EXT & $K_1K_2$/$YH$ & 1.24 & 1.04 &  1.7 \\
2015-05-01 & 95.C-0273 & IRDIS DPI   & B\_H    & 1.16 & 1.29 &  1.7 \\
\hline
\multicolumn{7}{c}{ALMA} \\
\hline
Observing date     & Prog. ID & Mode & Resolution &Frequency range & PWV      & Integration \\
$[$YYYY-MM-DD$]$   &          &      & $[$kHz$]$  &  $[$GHz$]$     & $[$mm$]$ &   $[$s$]$ \\
\hline
\multirow{2}{*}{2014-03-20}   & \multirow{2}{*}{2012.1.00437.S} & Continuum & 31250.00  &   $211.91-228.97$ & \multirow{2}{*}{0.79} & \multirow{2}{*}{120.96} \\
             &                & Gas       & 488.28 &   $230.05-230.99$ & & \\
\hline
\end{tabular}
\end{table*}

\begin{figure*}
\includegraphics[width=2.0\columnwidth]{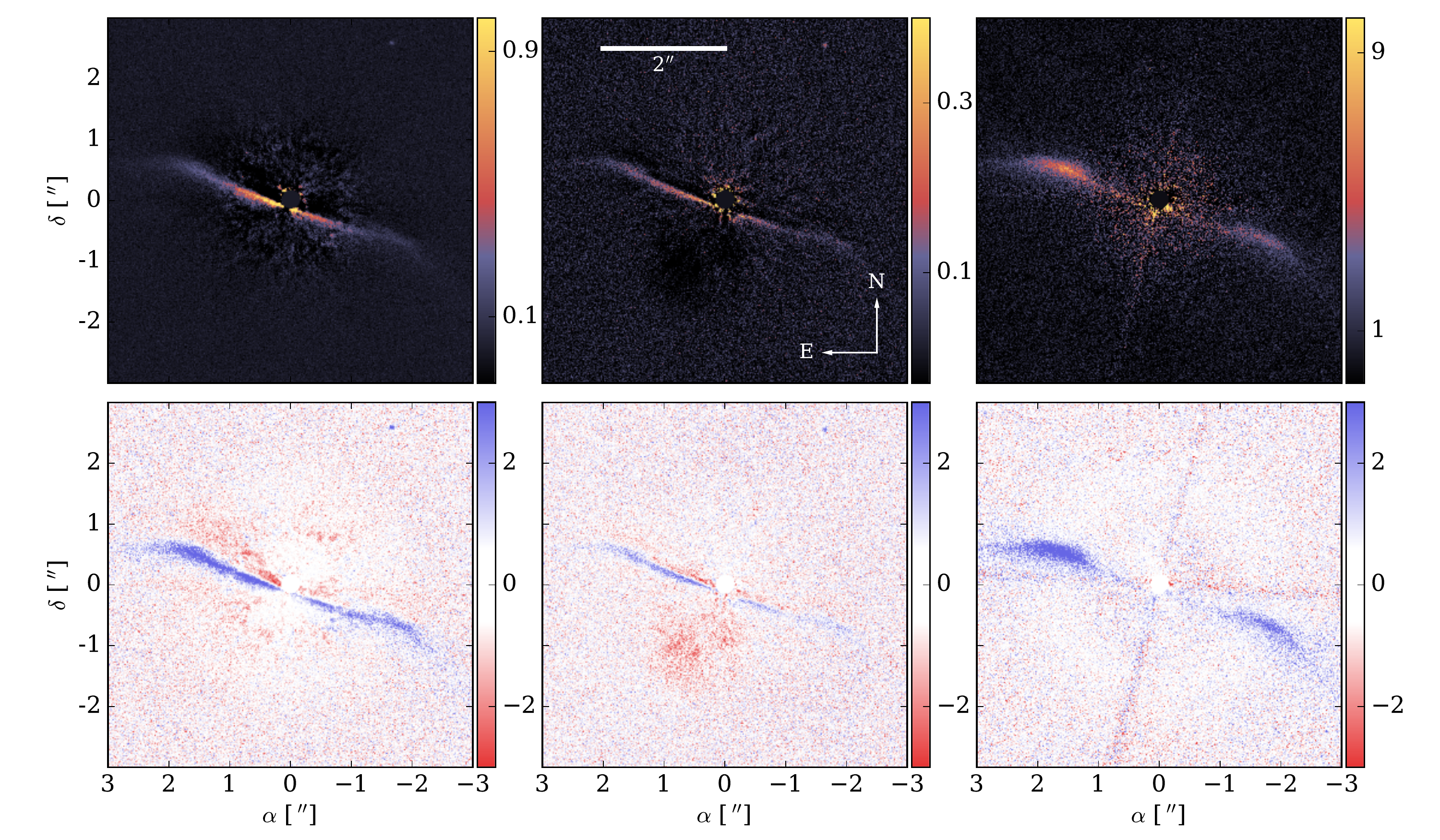}
\caption{Reduced SPHERE observations of HD\,61005 used in the analysis. North is up, East is left. From left to right; IRDIS ADI in $H$ and $K$ bands (PCA with 6 components), and IRDIS DPI $Q_{\phi}$ in $H$ band. Top row shows the data in linear stretch, with a central mask of $0.15^{\prime\prime}$, and the bottom row shows estimated signal-to-noise maps (see text for detail), with a stretch between [$-3\sigma$, $3\sigma$].}
\label{fig:all_data}
\end{figure*}

\subsection{VLT/SPHERE IRDIS observations and data reduction}

The star HD\,61005 was observed with the VLT/SPHERE (\citealp{Beuzit2008}), within the guaranteed time consortium. The observations were obtained in different instrumental set-ups in February, March, and May 2015, using the dual-band imager (IRDIS, \citealp{Dohlen2008,Vigan2010}), the integral field spectrograph (IFS, \citealp{Claudi2008}), and the dual-polarization imager (IRDIS DPI, \citealp{Langlois2014}). 

\subsubsection{IRDIS dual-band observations}

The February observations were conducted in the $H_2H_3$ dual band (centered on $1.59$ and $1.67$\,$\umu$m) for IRDIS and the $YJ$ band ($0.95 - 1.35$\,$\umu$m, at spectral resolution R$\sim54$) for IFS. The March observations used the $K_1K_2$ filters (centered on $2.11$ and $2.25$\,$\umu$m) for IRDIS and the $YH$ band ($0.95-1.65$\,$\umu$m, R$\sim33$) for IFS. All of these observations were performed using an apodized Lyot coronagraph, consisting of a focal mask with a diameter of $185$\,milli-arcsec (N\_ALC\_YJH\_S) and a corresponding pupil mask. Coronagraphic observations were performed in pupil stabilized mode to use angular differential imaging (ADI) post-processing (\citealp{Marois2006}) to attenuate residual speckle noise. The observation strategy can be summarized as follows: 1) Photometric calibration: imaging of star offset from coronagraph mask to obtain PSF for relative photometric calibration; 2) Centering: imaging with star behind mask with four artificially induced satellite spots for centering; 3) Science: coronagraphic sequence; 4) Centering: same as point two; 5) Photometric calibration: same as point one; 6) Sky background observation using same DIT as coronagraphic sequence. Finally, true north and plate scale are determined using astrometric calibrators as part of the SPHERE GTO survey for each run (\citealp{Maire2015}).

Basic reduction of the IRDIS data (background subtraction, flat fielding, centering) was performed using the SPHERE Data Reduction Handling (DRH) pipeline (\citealp{Pavlov2008}, version 15.0). The output consists of cubes for each filter, re-centered onto a common origin using the satellite spot reference. The cubes are then corrected for the true north position determined from the astrometric calibrations and for distortion. We collapsed the two filters together to increase the signal-to-noise ratio (S/N), and from now on we will refer to the $H_2H_3$ and $K_1K_2$ datasets as $H$ and $K$ observations, respectively. The datacubes in both bands were then processed using a principal component analysis (PCA, using the implementation of the \texttt{scikit-learn} Python package, \citealp{Pedregosa2011}) approach from all frames within the data-cube.

\subsubsection{IRDIS dual polarization observations}

On May 1, 2015, the target was observed using IRDIS in DPI mode in $H$ band. The same coronagraph was used for these observations. IRDIS DPI splits the light into two perpendicular polarization directions imaged at the same time on the same detector. Full cycles of half-wave plate (HWP) positions ($0$, $22.5$, $45$, and $67.5^{\circ}$) were taken to construct the Stokes $Q$ and $U$ vectors. The strategy of the observations was to take as long exposures as possible without saturating the detector just outside the coronagraph to achieve the best possible inner working angle as well as the best S/N for the outer part of the disk. Two integration times (DIT=$64$\,s with a total integration time of 768\,s and DIT=16\,s with a total integration time of 3008\,s) were used. IRDIS suffers from a de-polarization effect at certain detector position angles, which depend on the parallactic angle at the time of observation. Because the parallactic angle changed rapidly during the observations, we updated the detector angle at regular intervals.

The DPI data were reduced using a custom pipeline that is different to the DRH pipeline, which closely follows the processes described in \citet{Avenhaus2014}, using the double-difference method (see also \citealp{Canovas2011}) to construct Stokes $Q$ and $U$ vectors from the data. The pipeline has been adapted to suit the IRDIS instrument. The data were centered using the centering frames taken just before and after the science observations. A de-rotation was applied to bring all files to the same orientation (see section above). Furthermore, the frames were corrected to take account of the fact that the IRDIS pixel scale differs slightly ($\sim 0.6$\%) in the two principal detector directions. The files were then corrected for true north as determined for IRDIS.

Because scattered light in an optically thin (debris) disk is expected to be polarized perpendicular to the line between the star and the image point in question, we then construct local Stokes $Q$ and $U$ vectors, denoted $Q_{\phi}$ and $U_{\phi}$ (see also \citealp{Benisty2015}). In case of single scattering of the stellar light, $Q_{\phi}$ is expected to contain the disk signal, while $U_{\phi}$ is expected to contain no signal (with possible exceptions when the optical depth is large, \citealp{Canovas2015}) but noise on the same level as the $Q_{\phi}$ image and can serve as a noise estimator. $Q_{\phi}$ and $U_{\phi}$ can be calculated as:
\begin{equation}
\begin{split}
Q_{\phi} = +Q\,\mathrm{cos}(2 \Phi) + U\,\mathrm{sin}(2 \Phi)\\
U_{\phi} = -Q\,\mathrm{sin}(2 \Phi) + U\,\mathrm{cos}(2 \Phi),
\end{split}
\end{equation}
where $\phi$ refers to the azimuth in polar coordinates, and $\Phi$ is the position angle of the location of interest ($x$, $y$), with respect to the stellar location ($x_0$, $y_0$) as:

\begin{equation}
\Phi = \mathrm{arctan}\frac{x - x_0}{y - y_0} + \theta,
\end{equation}
where $\theta$ corrects for instrumental effects such as a small misalignment of the half-wave plate.

During the data reduction process, one HWP cycle equivalent to 256\,s of data (DIT = 16\,s) was taken out because the telescope had lost tracking for a short amount of time, rendering this data unusable. The result are two pairs of $Q_{\phi}$ and $U_{\phi}$ images, one for the DIT=16\,s and DIT=64\,s observations each. These were then combined with a weighted average to produce the final $Q_{\phi}$ and $U_{\phi}$ images.

\subsubsection{IFS observations}

The IFS data proved difficult to be properly reduced at the time of this analysis, mainly because of centering problems. Presenting and analyzing these observations will be postponed for a future study.

\subsubsection{Processed images}

Figure\,\ref{fig:all_data} shows the final reduction for our dataset (with a central mask of radius $0.15^{\prime\prime}$ ). The top row displays the IRDIS ADI data in $H$ and $K$ bands (left and middle panels, respectively), and the IRDIS DPI $Q_{\phi}$ image (right), all with a linear stretch. For each image, the bottom row shows a S/N map, with a linear stretch between [$-3\sigma$, $3\sigma$]. The noise map is calculated from the reduced images and represents the standard deviation in concentric annuli, centered on the star, with a constant width (2 pixels). We did not mask out the disk when computing the noise maps, hence the uncertainties might be slightly over-estimated. For the DPI observations, the noise map is computed from the $U_{\phi}$ image which does not seem to contain any signal from the disk (Fig.\,\ref{fig:uphi} shows the $U_{\phi}$ image with the same linear stretch as the $Q_{\phi}$ image in Fig.\ref{fig:all_data}).

We note that the disk is not homogeneously detected at high S/N. The east side is detected at larger S/N in all datasets and it appears brighter than the west side. Such asymmetry, already reported in \citet{Buenzli2010}, are further investigated in this paper. For the sake of simplicity, in the rest of the paper, we refer to the ADI and DPI datasets as ``scattered'' and ``polarized'' observations, respectively. Strictly speaking, this is not correct as polarized photons must have been scattered by dust grains.

\subsection{ALMA observations}

HD\,61005 was observed with ALMA in Band\,6 (PI: David Rodriguez, program 2012.1.00437.S), in the frequency range $211.97-230.99$\,GHz. The target was observed several times, with precipitable water vapor ranging from 5.19 to 0.79\,mm. We only kept the observations performed on the 20 of March 2014 for which the water vapor was minimum. Out of the four spectral windows, three were used to derive the continuum emission ($211.97-228.97$\,GHz, with a $31250$\,kHz resolution), while the last one was used to search for CO gas emission ($230.05-230.99$\,GHz, with a $488.28$\,kHz resolution), which was not detected. Data processing was performed within CASA using the standard scripts provided by the observatory. We only kept the spectral windows used for the continuum observations and averaged the complex visibilities along the 128 different spectral channels (while flagging points with negative weights). Figure\,\ref{fig:uvplane} shows the ($u$, $v$) plane coverage for the continuum observations, with minimum and maximum baselines of 11.8 and 334.9\,m, respectively. The reconstructed image (with so-called briggs weighting and pixel size of $0.13^{\prime\prime}$) is shown in the left panel of Fig.\,\ref{fig:ALMA} (sensitivity of $0.09$\,mJy/beam). The beam size is $1.36^{\prime\prime} \times 0.73^{\prime\prime}$ (48\,au$\times$26\,au) with a position angle of $-86.5^{\circ}$. We do not attempt to measure the total flux of the disk directly from these observations, but we do estimate it when modeling the complex visibilities (Section\,\ref{sec:alma}).

\subsection{Spectral energy distribution}

\begin{table}
\caption{Broadband photometric measurements of HD\,61005, and the equivalent widths of the far-IR filters (see text for details).\label{tab:flux}}
\centering
\begin{tabular}{@{}lcccr@{}}
\hline\hline
 $\lambda$ & $F_{\nu}$ & $\sigma$ & EW & Instrument \\
 $[\umu$m] & [mJy] & [mJy] & [$\umu$m] & \\
\hline
0.428  &   895.17 &  14.02 &        & TYCHO B    \\
0.534  &  1810.23 &  18.34 &        & TYCHO V    \\
1.235  &  2753.74 &  65.94 &        & 2MASS J    \\
1.662  &  2440.48 & 103.40 &        & 2MASS H    \\
2.159  &  1738.75 &  38.43 &        & 2MASS Ks   \\
3.353  &   819.28 &  31.69 &        & WISE W1    \\
4.603  &   453.05 &   8.76 &        & WISE W2    \\
11.56  &    78.40 &   1.08 &        & WISE W3    \\
22.09  &    44.28 &   1.55 &        & WISE W4    \\
68.92  &   717.00 &   5.33 &  21.41 & PACS Blue  \\
97.90  &   703.58 &   6.84 &  31.29 & PACS Green \\
153.94 &   472.65 &  14.58 &  69.76 & PACS Red   \\
251.50 &   235.6  &  13.5  &  67.61 & SPIRE PSW  \\
352.83 &   118.9  &   7.6  &  95.75 & SPIRE PMW  \\
511.60 &    49.8  &   5.2  & 185.67 & SPIRE PLW  \\
1300.0\tablefootmark{a} &     4.6  &   0.7  & 105.40 & ALMA Band\,6  \\
\hline
\end{tabular}
\tablefoottext{a}{Results from the modeling of the ALMA data (Section\,\ref{sec:alma}).}
\end{table}

The star HD\,61005 was observed by {\it Herschel} (\citealp{Pilbratt2010}) with the Photodetector Array Camera and Spectrometer instrument (\citealp[PACS,][]{Poglitsch2010}), within the program OT2\_tcurrie\_1. The observation numbers (OBSID) are the two pairs 1342270977, 1342270978 and 1342270979, 1342270980 for the 70\,$\umu$m and 100\,$\umu$m observations, respectively. The 160\,$\umu$m map used the four OBSID combined. The data were processed using the HIPE software (build 12.0.2083, \citealp{Ott2010}), the very same way as described in \citet{Olofsson2013}. HD\,61005 was also observed with the Spectral and Photometric Imaging Receiver instrument (\citealp[SPIRE,][]{Griffin2010}) in small scan map mode (OBSID: 1342255147 within the program OT2\_kstape01\_1). We used the Timeline Fitter task in HIPE to derive SPIRE photometry for our target. Calibration errors ($\sim$\,5.5\%, \citealp{Bendo2013}) are included in the uncertainties. We also gathered photometric observations using VOSA\footnote{http://svo2.cab.inta-csic.es/theory/vosa/} (\citealp{Bayo2008}) and the dataset used to build the SED can be found in Table\,\ref{tab:flux}. The meaning of the third column is explained in Section\,\ref{sec:sed}.

Finally, we downloaded the {\it Spitzer}/IRS spectrum from the Cornell Atlas of Spitzer/IRS Sources database\footnote{The Cornell Atlas of Spitzer/IRS Sources is a product of the Infrared Science Center at Cornell University, supported by NASA and JPL. http://cassis.sirtf.com/atlas/query.shtml} (\citealp{Lebouteiller2011}).

\subsection{Stellar parameters}

The stellar photospheric model is taken from the ATLAS9 Kurucz library \citep{Castelli1997} with an effective temperature of $T_{\star} = 5500$\,K (\citealp{Casagrande2011}). With the dilution factor used to scale the photospheric model to the optical and near-IR photometric measurements, at a distance of 35.4\,pc, we find a radius $R_{\star} = 0.84$\,$R_{\odot}$. We derived a luminosity of $L_{\star} = 0.58$\,$L_{\odot}$. To derive the stellar mass, which will become important when discussing the dust properties and the effect of radiation pressure on dust grains, we use isochrones from \citet{Siess2000}, for an age of 40\,Myr and effective temperature of $5500$\,K. We find that the stellar mass must be of about $1.1$\,$M_{\odot}$ (the corresponding luminosity matching our estimated $L_{\star}$). We find a slightly smaller mass ($1$\,$M_{\odot}$) when using the isochrones from \citet{Baraffe2015}, but the differences may arise from different model prescriptions (e.g. overshooting). In the following, we adopt a mass of $1.1$\,$M_{\odot}$. The SED with the broadband photometric measurements, the {\it Spitzer}/IRS spectrum as well as the photospheric model are shown in Fig.\,\ref{fig:SED} of Section\,\ref{sec:sed}.

\subsection{Preamble on the modeling strategy}

In this study, we aim to model observations from different facilities, at different wavelengths, using different techniques (interferometry and direct imaging). Therefore, before detailing the modeling strategy for each individual dataset, we provide a quick preamble on the methodology.

We first model the ALMA data (Section\,\ref{sec:alma}) assuming a circular disk, fitting the reference radius ($r_0$, where the dust density peaks), the outer slope for the dust density distribution ($\alpha_{\mathrm{out}}$), the position angle ($\phi$), the inclination ($i$), and the total flux at 1.3\,mm ($f_{\mathrm{1300}}$).

Prior to the modeling of the SPHERE observations, we attempt to constrain some of the dust properties, to limit the number of free parameters. We use the best fit results for the inclination and position angle from the modeling of the ALMA data to derive the polarized intensity as a function of the azimuthal angle from the $Q_{\phi}$ image. We constrain the minimum and maximum grain sizes ($s_{\mathrm{min}}$ and $s_{\mathrm{max}}$, respectively) as well as the porosity fraction of the dust grains (Section \ref{sec:pfunc}). This enables us to reduce the pool of free parameters when modeling the SPHERE DPI observations. For the SPHERE ADI observations, we use the Henyey-Greenstein approximation for the phase function, which disconnects the modeling process from the aforementioned dust properties. Thanks to the great complementarity between the ADI and DPI observations, we model them simultaneously to best constrain the azimuthal and radial dust density distribution (Section\,\ref{sec:sphere}).

Finally, in Section\,\ref{sec:sed}, we model the SED of HD\,61005, using the results inferred from the modeling of the SPHERE data on the location of the disk to derive stringent constraints on the dust properties (minimum grain size, dust composition, and total dust mass).


\section{The parent planetesimal belt: constraints from the ALMA observations}\label{sec:alma}

Roughly speaking, different wavelengths trace different grain sizes. Therefore, we chose to model the ALMA observations independently of the SPHERE ones, the latter probing the small dust grains in the debris disk while the millimeter observations most likely trace a population of larger grains that more closely follow the parent planetesimals' belt.

\subsection{Modeling strategy}

The modeling of the ALMA observations is performed in the Fourier space, attempting to reproduce both the real and imaginary parts of the complex visibilities (averaged along the spectral dimension). In Appendix\,\ref{app:DDIT}, we explain how we generate synthetic images and the different notations are summarized in Table\,\ref{tab:notations}. From a synthetic image at the wavelength of $1.3$\,mm, we first scale the total flux of the image to the free parameter $f_{1300}$ (in mJy) before computing the Fourier transform of the image. We then interpolate the Fourier transform at the spatial frequencies of the observations. The goodness of fit is the sum of the weights (estimated in CASA\footnote{The absolute values of the weights derived within CASA may be inaccurate but their relative values are not.}) times the squared difference between the observed and modeled complex visibilities (the weights are proportional to $1 / \sigma^2$). We consider the following free parameters: the inclination $i$, the position angle $\phi$, the total flux of the disk at $1.3$\,mm $f_{1300}$, the reference radius\footnote{Here we assume the disk is circular.} $r_0$, and the outer power-law slope for the dust distribution $\alpha_{\mathrm{out}}$, which is parametrized as
\begin{equation}\label{eqn:dustdens}
n \propto \left[ \left( \frac{r}{r_0} \right)^{-2\alpha_{\mathrm{in}}} + \left( \frac{r}{r_0} \right)^{-2\alpha_{\mathrm{out}}} \right] ^{-1/2},
\end{equation}
where $r$ is the distance from the star and $n$ the number density. The inner power-law slope is set to $\alpha_{\mathrm{in}} = 5$, preliminary tests indicating this parameter is poorly constrained by the observations. To find the most probable solution, we use an affine invariant ensemble sampler Monte-Carlo Markov Chain, implemented in the \texttt{emcee} package, using $200$ walkers, a burn-in phase of $500$ iterations and a total length of the chains of $2000$ iterations after the burn-in phase. At the end of the run, we find that the mean acceptance fraction (the mean fraction of steps accepted for each walker within the chain) is of $0.48$ (a good sign of convergence and stability, \citealp{Gelman1992}). The maximum auto-correlation time for all the parameters is of 60 steps, indicating that the chains should have stabilized by the end of the simulations.

\subsection{Results}\label{sec:res_alma}

\begin{table}
\centering
\caption{Best fit results for the modeling of the ALMA observations.\label{tab:alma}}
\begin{tabular}{@{}lccc@{}}
\hline\hline
Parameter & Uniform prior & $\sigma_{\mathrm{kde}}$ & Best-fit value \\
\hline
$r_0$ [au]              & $[40, 80]$     & $0.2$  & $66.4_{-8.7}^{+6.1}$ \\
$\alpha_{\mathrm{out}}$ & $[-15, -1.5]$  & $0.1$  & $-6.6_{-6.1}^{+1.4}$ \\
$\phi$ [$^{\circ}$]     & $[55, 85]$     & $0.1$  & $70.7_{-2.3}^{+1.9}$ \\
$i$ [$^{\circ}$]        & $[79, 89]$     & $0.1$  & $84.5_{-2.5}^{+2.9}$ \\
$f_{1300}$ [mJy]        & $[1, 20]$      & $0.1$  & $4.6_{-0.6}^{+0.7}$  \\
\hline
\end{tabular}
\end{table}

\begin{figure*}
\includegraphics[width=2\columnwidth]{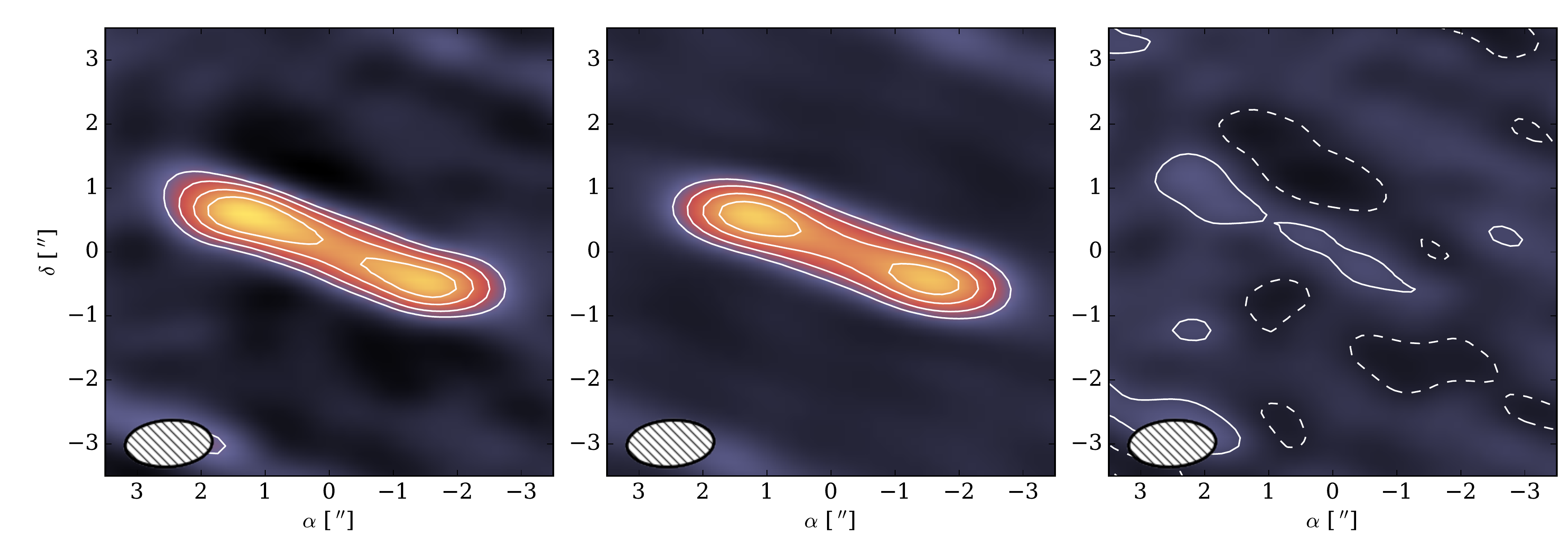}
\caption{From left to right: observations, best-fit model and residuals of the ALMA data. For all panels the color map is a linear stretch between $-0.27$ and $1.23$\,mJy/beam. The standard deviation estimated in an empty region of the observations is of $0.09$\,mJy/beam. For both the observations and the model the contours are set at [$3$, $5$, $7.5$, $10$]$\sigma$, and [$-\sigma$, $\sigma$] for the residuals (no residuals beyond the $2\sigma$ level). The beam size is shown in the lower left corner of each panels.}
\label{fig:ALMA}
\end{figure*}

\begin{figure}
\includegraphics[width=\columnwidth]{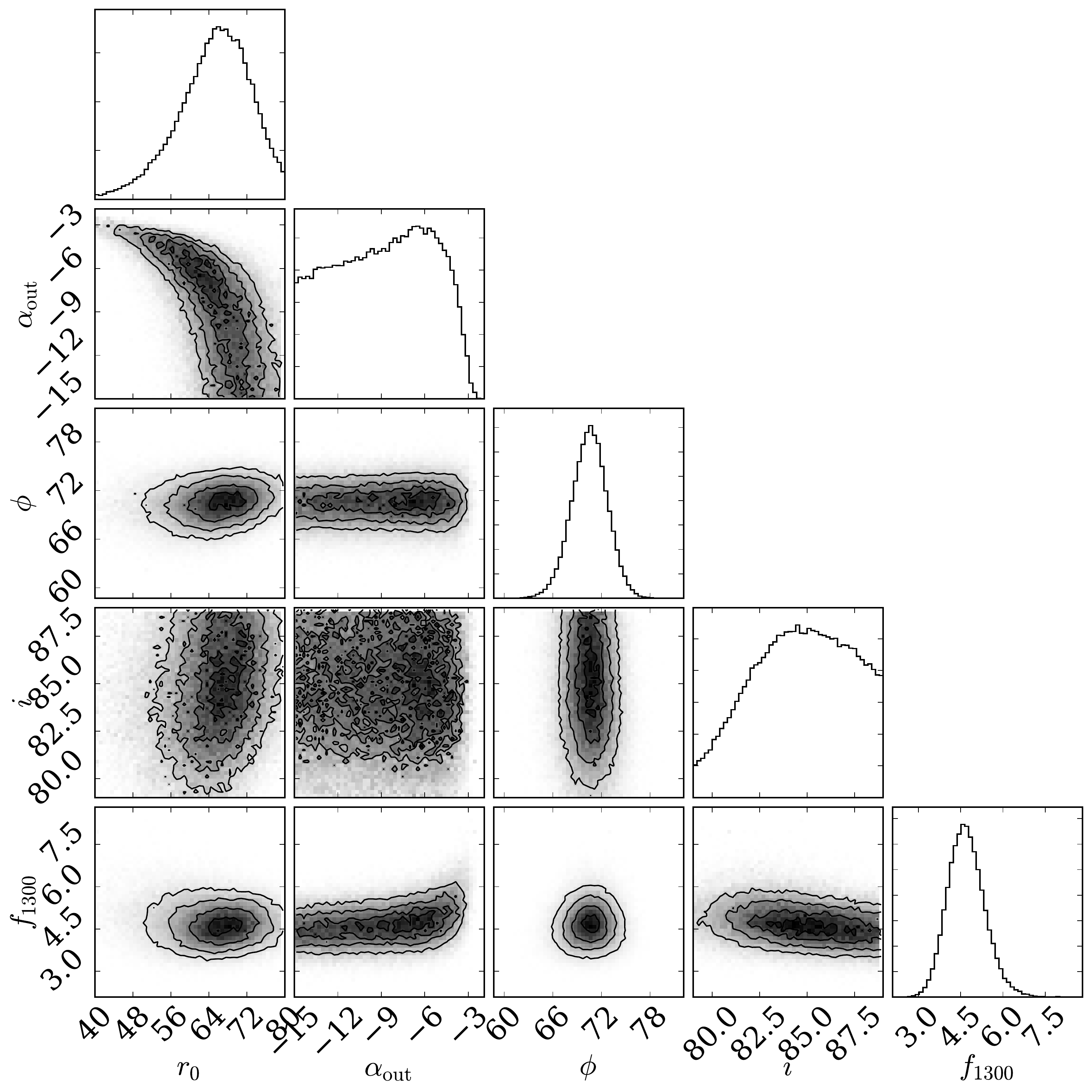}
\caption{One- and two-dimensional (diagonal and lower triangle, respectively) projections of the posterior probability distributions for the results of the modeling of the ALMA observations.}
\label{fig:emcee_ALMA}
\end{figure}

The projected posterior probability distributions are displayed in Fig.\,\ref{fig:emcee_ALMA}, for the different free parameters (using the \texttt{triangle} Python package, \citealp{ForemanMackey2014}). To derive the best-fit values as well as the uncertainties, we smooth the distributions with a kernel density estimator (the width of the Gaussian kernel $\sigma_{\mathrm{kde}}$ are reported in Table\,\ref{tab:alma}), and the best-fit value is the peak position of the distribution. The confidence intervals ($a_1$, $a_2$) for the parameter $a$ are estimated as follows:
\begin{equation}
\int_{a_{\mathrm{min}}}^{a_1} p(a) da = \int_{a_2}^{a_{\mathrm{max}}} p(a) da = \frac{1 - \gamma}{2},
\end{equation}
where $\gamma = 0.68$ and $p(a)$ is the smoothed posterior probability distribution (integral normalized to $1$) for parameter $a$ (e.g., \citealp{Pinte2008}). We note that not all the distributions reach zero on each side of their maximum (especially for $\alpha_{\mathrm{out}}$ and $i$) and, therefore, these uncertainties should be treated carefully. Table\,\ref{tab:alma} summarizes our results for the  modeling of the Band 6 observations, and Fig.\,\ref{fig:ALMA} shows the observations, the best-fit model, and the residuals (from left to right). The synthetic image of the best fit model is processed through CASA (using the \texttt{ft} method with the same antenna configuration as the observations) and the image is reconstructed with the \texttt{clean} algorithm with the same parameters as for the observations. We note that there are some residuals on the east side that may suggest that the disk is brighter on one side, even at mm wavelength. However, these residuals are below $3\,\sigma$, therefore we cannot conclude they are significant. The apparent brightness asymmetry could be due to the asymmetric ($u$, $v$) coverage of the observations.

Overall, we find that most of the parameters are well constrained, except for the outer power-law slope of the dust density distribution, for which we can safely exclude slopes shallower than $\alpha_{\mathrm{out}} = -4$. This is explained by the beam size of the observations which is larger than the debris disk for steep values of $\alpha_{\mathrm{out}}$. Otherwise, we find the reference radius of the disk to be $r_0 \sim 66$\,au, the position angle $\phi \sim 70.7^{\circ}$, the inclination $i \sim 84.5^{\circ}$, and the flux at 1.3\,mm $f_{1300} \sim 4.6$\,mJy. These results agree well with the parameters reported in \citet[][$i = 84.3\pm1^{\circ}$, $\phi = 70.3\pm1^{\circ}$, $r_0 = 61.25\pm0.85$\,au]{Buenzli2010} and \citet[][$r_0 = 67\pm2$\,au, $\phi = 71.5\pm5^{\circ}$]{Ricarte2013}. The relatively large beam size of the ALMA observations can explain the slight discrepancy for $r_0$ between the modeling of the ALMA data and the value inferred by \citet{Buenzli2010}. This value will be revisited when modeling the SPHERE observations (Section\,\ref{sec:sphere}). \citet{Ricarte2013} obtained a total flux of $7.2\pm0.3$\,mJy at $1.3$\,mm with their SMA observations (\citealp{Steele2016} obtained $8.0 \pm 0.8$\,mJy analyzing the same observations), while we find the total flux to be well constrained at $4.6\pm0.7$\,mJy (within the SMA and ALMA respective $3\sigma$ uncertainties). The shortest baselines being of $B_{\mathrm{min}} \sim 11$ and $16$\,m (for the ALMA and SMA observations, respectively), the largest scales the observations are sensitive to are of the order of $14.2^{\prime\prime}$ and $10.1^{\prime\prime}$, respectively ($0.6 \lambda / B_{\mathrm{min}}$), much bigger than the disk. It is therefore unlikely that flux from the disk is filtered out by the interferometers. The differences between the SMA and ALMA data may arise from missing frequencies, differences in beam sizes, or calibration uncertainties (the uncertaintites reported by \citealp{Steele2016} being more conservative than the ones reported by \citealp{Ricarte2013}).

Finally, we note that the famous wings responsible for the disk's nickname are not detected in the ALMA data. Despite a good angular resolution, a disk model (without wings) can successfully reproduce the observations, and we see no trace of the wings in the residuals (with an rms of $\sim 0.09$\,mJy/beam). Our results therefore agree with the ones of \citet{Ricarte2013}; only small dust grains are likely present in the wings.

\section{Constraining the dust radial distribution from the SPHERE observations}\label{sec:sphere}

We model the SPHERE images by producing synthetic images at the central wavelength of the $H$-band observations, $\lambda_{\mathrm{c}} = 1.63$\,$\umu$m. Given the low S/N of the $K$-band ADI observations, preliminary attempts to model these data showed that the dust distribution cannot be better constrained than with the $H$-band observations. We therefore focus the modeling effort on the $H$-band ADI and DPI data.

Figure\,\ref{fig:all_data} highlights the complementarity of both the scattered and polarized light images; the ADI and DPI data have very different S/N at the ansae and along the semi-minor axis of the disk. Combining both datasets, therefore, offers the opportunity to study the dust distribution in great detail.

The modeling process has a high dimensionality with many possible free parameters, and regions of intermediate to low S/N. Therefore, to obtain novel yet reliable constraints on the dust distribution we choose to perform a prior analysis on the observations to reduce the number of free parameters. Having a proper description of the polarized phase function prior to the modeling of the DPI observations greatly helps reducing the dimensionality of the modeling (e.g., the minimum and maximum grain sizes as well as the porosity of the dust grains). In this section, we first describe how the polarized phase function is derived and modeled, then we present the modeling strategy and summarize the results we obtain.

\subsection{Phase function of the polarized light}\label{sec:pfunc}

To compute the polarized phase function from the DPI dataset, we define an elliptical mask with the following parameters: inner and outer radii ($r_{\mathrm{in}}$ and $r_{\mathrm{out}}$), the inclination $i$, and the position angle $\phi$. For each pixel within the elliptical mask, we compute the scattering angle as the dot product of the line of sight and the location of the pixel with respect to the star. We divide the elliptical mask in two, for the east and west sides and, for each side, we compute the minimum and maximum scattering angles. We then divide each side of the mask into $30$ smaller regions corresponding to different bins of the phase function (see Fig.\,\ref{fig:ell_pfunc} for an illustration). Each pixel is multiplied by its squared distance to the star, to account for illumination effect. The measured phase function is found by averaging the flux in the observations, in each individual region. Since the observed uncertainties $\sigma_i$ are not the same for each pixel within a given intersection, the ``average'' uncertainty in the intersection is computed as follows:
\begin{equation}\label{eqn:uncertainties}
\sigma = \left( \sum_{i=1}^{N} \frac{1}{\sigma_i^2} \right)^{-1/2},
\end{equation}
where $N$ is the number of pixels in the considered region. The phase function is then normalized to its maximum value (east and west sides are divided by the same value).

\begin{figure}
\includegraphics[width=\columnwidth]{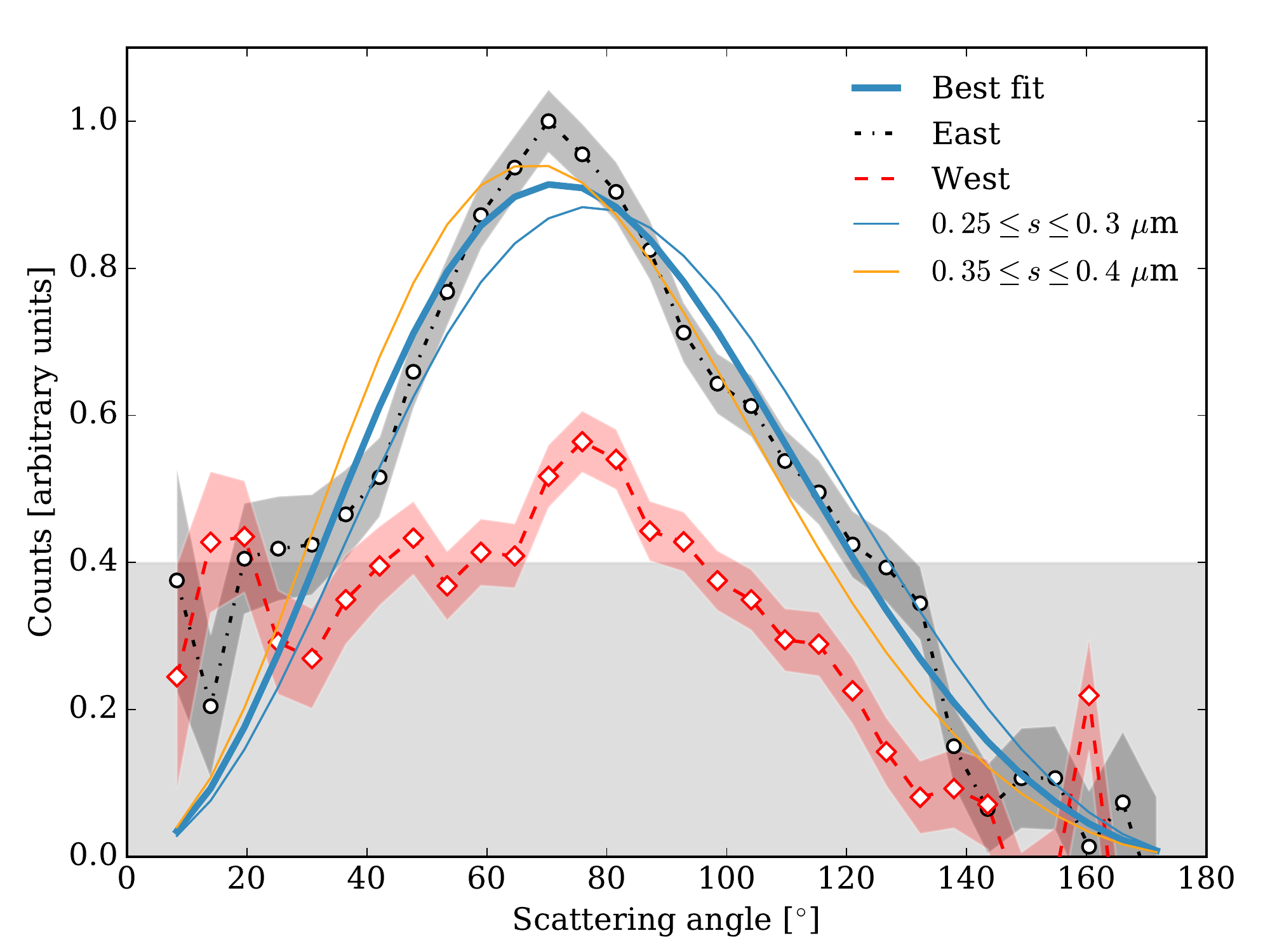}
\caption{Azimuthal dependency of the polarized intensity in the DPI $H$-band observations, in black and red for the East and West sides, respectively. The shaded curves are representative of the uncertainties estimated from the noise map and the shaded gray area denotes low S/N regions estimated from the S/N map in Fig.\,\ref{fig:all_data}. The thin solid lines display examples for different grain sizes for the same dust composition as the best fit.}
\label{fig:pfunc_dpi}
\end{figure}

For the elliptical mask, we use the best-fit results of the modeling of the ALMA observations (see Table\,\ref{tab:alma}) for $i$ and $\phi$ and choose $r_{\mathrm{in}}$ and $r_{\mathrm{out}}$ small and large enough ($50$ and $72$\,au, respectively) so that they encompass the trace of the debris disk. Figure\,\ref{fig:pfunc_dpi} shows the phase function for the DPI $H$-band observations, the east side being in black, and the west side in red. The uncertainties are shown in a shaded color, and the gray area indicates regions of low S/N, which were estimated from the S/N map derived from the $U_{\phi}$ data. As can be seen in Fig\,\ref{fig:all_data} (bottom right panel), the disk is only detected for a narrow range of azimuthal angles on the west side. This strongly suggests that we underestimated the uncertainties calculated using Eq.\ref{eqn:uncertainties}.

The difference in brightness between the two sides of the disk is striking in Fig.\,\ref{fig:pfunc_dpi} and the east side appears almost twice as bright as the west side for most of the scattering angles. Even though the S/N for the west side is relatively low it seems that the polarized phase function peaks at an angle compatible with the east side.

To alleviate the number of free parameters, we aim to constrain the grain size distribution for the modeling of the DPI observations directly from the phase function displayed in Fig.\,\ref{fig:pfunc_dpi}. We emphasise that we do not have access to the polarization degree ($\sqrt(Q^2 + U^2) / I$) as we do not have an unbiased measure of the total intensity $I$ prior to the modeling. Indeed, the ADI process introduces self-subtraction effects (e.g., \citealp{Milli2012}), which can eventually be quantified with a model that describes the observations well (which we do not yet have). Since the west side suffers from low S/N, we perform the modeling on the east side's phase function. To reproduce the phase function, we assume that the signal in the $Q_{\phi}$ image is proportional to the size dependent $S_{12}(s)$ element of the M\"uller matrix. The matrix enables us to compute the Stokes vectors $I$ and $Q$ for the scattered and polarized light, respectively. Assuming single scattering event (reasonable assumption in low density environment such as debris disks), the scattered light will be the product of the first diagonal element of the matrix $S_{11}$ times the stellar intensity $I_0$. The polarized intensity will be proportional to the second element of the first column of the matrix $S_{12}$ times $I_0$. We compute $S_{12}$ for different grain sizes between $s_{\mathrm{min}}$ and $s_{\mathrm{max}}$, using the Mie theory, and we average $S_{12}$ over a grain size distribution with a slope $p < 0$ (d$n(s) \propto s^p$d$s$) as follows,
\begin{equation}
    S_{12}^{\mathrm{avg}} = \cfrac{\int_{s_{\mathrm{min}}}^{s_{\mathrm{max}}} S_{12}(s) \times s^p \mathrm{d}s}{\int_{s_{\mathrm{min}}}^{s_{\mathrm{max}}} s^p \mathrm{d}s}.
\end{equation}
We then compute the best scaling factor $f_{S_{12}}$ (to be multiplied to $S_{12}^{\mathrm{avg}}$) that will minimize the difference between the profiles $S_{12}^{\mathrm{avg}}$ and $S_{12}^{\mathrm{obs}}$ (with uncertainties $\sigma$)
\begin{equation}\label{eqn:minimize}
    f_{S_{12}} = \cfrac{\sum \left( \cfrac{S_{12}^{\mathrm{obs}} \times S_{12}^{\mathrm{avg}}}{\sigma^2} \right)}{\sum \left(\cfrac{S_{12}^{\mathrm{avg}}}{\sigma}\right)^2}.
\end{equation}
We perform a simple grid search over the following parameters: $s_{\mathrm{min}}$, $s_{\mathrm{max}}$, and the optical properties of the dust grains. We fix $p = -3.5$, as expected for a collisional cascade in a debris disk (\citealp{Dohnanyi1969}). For the dust composition, we consider a base medium of amorphous silicate with olivine stoichiometry (MgFeSiO$_4$, \citealp{Dorschner1995}) to which we can add some porosity using the Bruggeman mixing rule. The minimum grain size can vary between $0.01$ and $10$\,$\umu$m. To constrain the maximum grain size, we vary the quantity  $\Delta s$ ($= s_{\mathrm{max}} - s_{\mathrm{min}}$) between $0.01$ and $100$\,$\umu$m (in log space), and finally the porosity can change by steps of $10$\%. We find that the polarized phase function at $1.63$\,$\umu$m is best reproduced by small spherical dust grains,
with typical sizes in  the range $0.3 \leq s \leq 0.35$\,$\umu$m and a porosity fraction of $\sim 80$\%. The best-fit solution is shown with a thick cyan line in Fig.\,\ref{fig:pfunc_dpi}, and reproduces well both the overall shape and the peak position of the observed phase function. Also shown in Fig.\,\ref{fig:pfunc_dpi} are two examples for different grain size distributions around $0.25-0.3$\,$\umu$m and $0.35-0.4$\,$\umu$m, to illustrate how sensitive the phase function is with respect to the grain sizes. We note that, within such a narrow range of sizes, the value of the slope $p$ of the grain size distribution remains unconstrained. In this section, we assumed that the polarized flux is directly proportional to $S_{12}$, while it is also related to the dust density in the disk. In the rest of this paper, we try to determine the azimuthal distribution of the dust in the disk. Therefore, this prior analysis must be regarded as a first order approximation. The motivation of this prior analysis was to reduce the dimensionality of the modeling, but it also comes at a slight cost in the interpretation of the grain size distribution. The main conclusion of this section is that it does not seem that we observed large dust grains (which would have a stronger $S_{12}$ signal at small scattering angles). In Section\,\ref{discuss:dust} we discuss this result further, but it could also be the consequence of low S/N along the semi-minor axis of the disk.

\subsection{Modeling strategy}

Since both the ADI and DPI datasets were taken at the same wavelength, for a given set of parameters we compute two images: an unpolarized light image using the Henyey-Greenstein (HG) analytical prescription of phase function $S_{11}$ (\citealp{Henyey1941}) and a polarized light image using the Mie theory. Using the HG approximation, which is parametrized by the anisotropic scattering factor $g$ ($-1 \leq g \leq 1$), gives us more control when trying to reproduce the observed phase function (hence less free parameters to be considered during the modeling). To reproduce the DPI observations, we use the grain properties derived previously (Section\,\ref{sec:pfunc}). The absorption and scattering efficiencies, as well as the M\"uller matrix element $S_{12}$, are computed with the Mie theory, which is valid for compact spherical grains. This approach may not appear self-consistent, but it is a way to disentangle the modeling of unpolarized and polarized light that may not be well accounted for by spherical grains (e.g., \citealp{Milli2015}).

The pool of free parameters includes the reference radius $r_0$, the inclination $i$, the position angle $\phi$, the opening angle $\psi$, and the outer power-law slope $\alpha_{\mathrm{out}}$ for the dust density distribution. Preliminary tests indicated that we can hardly constrain the inner power-law slope, so we fixed $\alpha_{\mathrm{in}} = 5$. This enables us to focus on other parameters that describe the geometry of the debris disk. For instance, \citet{Buenzli2010} conclude that the eccentricity of the disk was not enough to explain the brightness asymmetry between the east and the west sides, and we aim to address this interpretation of previous observations. Therefore, we include the eccentricity $e$, the rotation angle $\Phi_{\mathrm{e}}$, the density damping $\eta$, and its azimuthal shape (via the width $w$ of the Gaussian profile) and its reference angle $\Phi_{\mathrm{\eta}}$ (see App.\,\ref{app:DDIT}). The azimuthal profile has the shape of a Gaussian profile with a $\sigma = w$, a peak of $1$ for the azimuthal angle $\Phi_{\eta}$ and a minimum value of $\eta \geq 0$ (hence an amplitude of $1-\eta$). 

Because we now consider azimuthal variations for the dust density distribution, it is parametrized slightly differently. The semi-major axis of the disk is defined as $r_0 / (1 - e^2)$. Although the formal denomination of $r_0$ is the so-called semi latus rectum, we will simply refer to it as the reference radius in the rest of the analysis. The radius at which the dust density peaks now depends on the azimuthal angle $\theta$. This radius $r_\theta$ is defined as $r_0 / (1 + e \times \mathrm{cos}\theta)$, and the azimuth-dependent dust density distribution $n(\theta)$ is parametrized similarly to Eq.\,\ref{eqn:dustdens}, replacing $r_0$ with $r_\theta$ for each azimuthal angle.

The dust mass $M_{\mathrm{dust}}$ is not varied, but the synthetic images are scaled during the fitting process. For the polarized images, the modeled image is the absolute value of the Stokes $Q_{\phi}$ parameter (we are assuming that there are no multiple scattering events in the debris disk). We scale the modeled image by a factor $f_{\mathrm{DPI}}$, which is found analytically to minimize the residuals (similarly to Eq.\ref{eqn:minimize}). For the ADI dataset, this approach is not possible because the post-processing of the data cube can introduce self-subtraction effects. We therefore opted for a forward modeling strategy, similar to the one described in \citet{Thalmann2014}. For a given set of parameters, we compute one synthetic image at the wavelength $1.63$\,$\umu$m. We then produce a cube of 64 images, each one rotated to match the parallactic angles of each frame of the observations (total rotation of $93.3^{\circ}$). Each frame of the cube of synthetic images is multiplied by $f_{\mathrm{ADI}}$ and is then subtracted from the corresponding observed frame. The PCA process is performed, keeping only the six main components. Another approach would be to perform the PCA on the observations, save the coefficients, and apply them to the modeled cube. But the observations contain signal from the disk that may be accounted for in some of the principal components (even though they should be similarly subtracted in the modeled cube). Therefore, to ensure as little subtraction as possible, we chose to perform the PCA on the model-subtracted cube. The end goal being to minimize the flux in the final image. 

The goodness of the fit is the sum of the squared ratio between the final images (residuals for the scattered and polarized data) and the noise map (bottom row of Fig.\,\ref{fig:all_data}). For the modeling of both the ADI and DPI observations, we therefore have a total of 12 free parameters: $r_0$, $i$, $\alpha_{\mathrm{out}}$, $e$, $\Phi_{\mathrm{e}}$, $\phi$, $\eta$, $\Phi_{\mathrm{\eta}}$, $w$, $\psi$, $g$, and $f_{\mathrm{ADI}}$. Here, we also use the \texttt{emcee} Python package, with 100 so-called walkers in the chain, and first burn in 500 runs for each of these walkers. We then run the chain for 2\,000 iterations in total (see \citealp{MillarBlanchaer2015} for a similar modeling approach). To speed up the process, we cropped and re-sampled the SPHERE images. The size of one pixel in the new image (or cube) is resampled to be $2.25$ times bigger than in the original images, the size of the new image being $200 \times 200$ pixels, and we use a central mask of radius $0.15^{\prime\prime}$. We chose not to convolve the synthetic images by a PSF because the cropped and down-sampled images have a pixel size of $0.028^{\prime\prime}$, while the approximation of the instrumental PSF with a Gaussian profile would have a width of $0.024^{\prime\prime}$ (approximating the PSF as an Airy disk for an 8\,m telescope). At the end of the run, we find that the mean acceptance fraction (the mean fraction of steps accepted for each walker within the chain) is of $0.35$, with a maximum auto-correlation time of 79 steps.

\subsection{Results}

\begin{figure*}[!htbp]
\includegraphics[width=2\columnwidth]{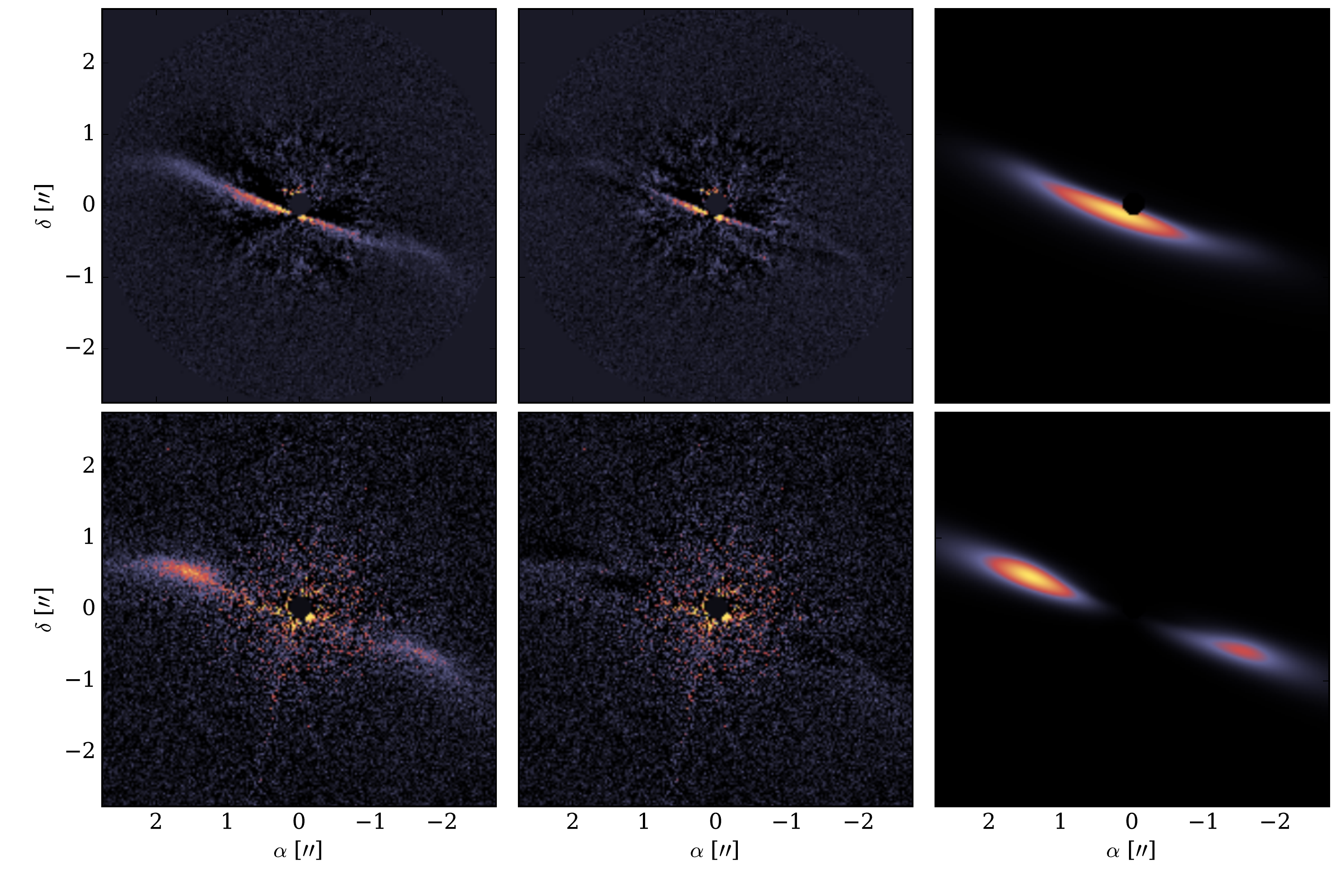}
\caption{\textit{Left to right}: observed, residuals, and best-fit models for the ADI and DPI datasets (top and bottom, respectively).}
\label{fig:residuals_H}
\end{figure*}

\begin{table}
\centering
\caption{Best fit results for the ADI and DPI $H$-band observations.\label{tab:res_H}}
\begin{tabular}{@{}lccc@{}}
\hline\hline
Parameter & Uniform prior & $\sigma_{\mathrm{kde}}$ & Best-fit value \\
\hline
$r_0$ [au]                       & $[40, 80]$     & $0.1$   & $60.4_{-0.5}^{+0.8}$ \\
$i$ [$^{\circ}$]                 & $[75, 88]$     & $0.1$   & $84.1_{-0.2}^{+0.2}$ \\
$\alpha_{\mathrm{out}}$          & $[-10, -1.75]$ & $0.01$  & $-2.70_{-0.1}^{+0.1}$ \\
$e$                              & $[0, 0.6]$     & $0.0025$& $0.093_{-0.014}^{+0.018}$ \\
$\Phi_{\mathrm{e}}$ [$^{\circ}$]    & $[0, 360]$     & $1.$    & $127.3_{-4.3}^{+12.7}$ \\
$\phi$ [$^{\circ}$]              & $[65, 75]$     & $0.1$   & $70.6_{-0.3}^{+0.2}$ \\
$\eta$                           & $[0, 1]$       & $0.0025$& $0.47_{-0.04}^{+0.02}$ \\
$\Phi_{\mathrm{\eta}}$ [$^{\circ}$] & $[0, 360]$     & $1.0$   & $138.1_{-5.6}^{+7.2}$ \\
$w$ [$^{\circ}$]                 & $[5, 180]$     & $1.0$   & $51.7_{-3.2}^{+8.8}$ \\
$\psi$                           & $[0.02, 0.12]$ & $0.005$ & $0.058_{-0.002}^{+0.001}$ \\
$g$                              & $[0, 0.95]$    & $0.001$ & $0.54_{-0.02}^{+0.01}$ \\
$f_{\mathrm{ADI}}$               & $[3, 8.5]$     & $0.01$  & $6.20_{-0.07}^{+0.09}$ \\
\hline
\end{tabular}
\end{table}

The projected posterior probability distributions are displayed in Fig.\,\ref{fig:pdf_H}, for the different free parameters. To derive the best-fit values as well as the uncertainties, we proceed similarly as in Section\,\ref{sec:res_alma}. The results are presented in Table\,\ref{tab:res_H} and Fig.\,\ref{fig:residuals_H} shows the observations, the residuals and the best-fit models (from left to right) for the ADI and the DPI data (top and bottom rows, respectively).

All parameters seem to be well constrained. The most probable solution has a semi-major axis of 60.9\,au ($r_0 / [1 - e^2]$) for an eccentricity of $0.093$. With a rotation angle of $\Phi_{\mathrm{e}} \sim 127^{\circ}$ the pericenter is located slightly toward the observer, on the east side\footnote{For $\Phi_{\mathrm{e}} = 180^{\circ}$, the pericenter would be along the semi-major axis on the east side. Smaller angles would move the pericenter towards the observer, while larger angles would move the pericenter towards the back side of the disk.}. An azimuthal density variation seems to be necessary to reproduce the observations with a damping factor $\eta$ of $\sim 0.47$ with a reference angle of $\sim 138^{\circ}$ (hence almost co-located with the pericenter of the eccentric disk). The azimuthal variation of the dust density distribution is a Gaussian profile with a width of $\sim 50^{\circ}$. The position angle and inclination are consistent with the results from \citet{Buenzli2010}. The aspect ratio of $\psi \sim h/r \sim 0.06$ agrees well with numerical simulations of the vertical structures of debris disks (e.g., \citealp{Thebault2009}). Finally, the phase function is anisotropic for low scattering angles with $g \sim 0.54$. The dust density distribution for the best-fit model, viewed from above the disk, is displayed in Fig.\,\ref{fig:best_sketch}.

Nonetheless, as shown in the residuals of the ADI observations, our best-fit model does not manage to remove all the signal along the semi-minor axis of the disk. It successfully suppresses most of the signal at larger scattering angles, but the best-fit model seems to fail at properly describing the scattering at smaller angles. We tried to implement a phase function with two weighted HG functions, but could not significantly improve the residuals. We included all the ``basic'' parameters related to the geometry of the disk ($i$, $\phi$, $r_0$, $\psi$) and yet failed to perfectly match the observations. Possible explanations can be related to the phase function (the HG phase function remains an approximation), the radial segregation of the grain size distribution, or the azimuthal dust density distribution (this will be further discussed in Section\,\ref{discuss:dust}). We crudely assumed a Gaussian profile for the azimuthal distribution, but the actual distribution could be skewed in one or another direction which could explain the brighter region along the semi-minor axis. Another (highly speculative but interesting) explanation could be that we may be seeing an inner disk. Because of the high inclination of the disk, an inner dust belt may appear to the observer as if it was merging with the main belt. The main challenge with this explanation is that we see no indication of this type of a belt in the DPI observations, but the noise is larger in the innermost regions of this dataset. One possible way to address this point would be to detect gas, which could trace velocities compatible with a radius smaller than $61$\,au. Yet no CO was detected in the ALMA observations (Section\,\ref{sec:CO}).

\section{Constraining the dust mineralogy from the SED}\label{sec:sed}

The SED of the debris disk around HD\,61005 is constructed from the fluxes reported in Table\,\ref{tab:flux} and the {\it Spitzer}/IRS spectrum. Modeling an SED from unresolved observations is a degenerate problem. We are basically trying to find the adequate temperature of the grains, which can be changed either by their radial distances, their sizes, or their nature. The modeling of the SPHERE images provide strong constraints on the geometric parameters of the disk, and we can therefore focus on better constraining the dust properties.

\subsection{Modeling the thermal emission from the disk}

To model the SED of HD\,61005, for a given set of parameters, the goodness of fit is computed as
\begin{equation}\label{eqn:chi2}
\chi^2 = \sum_i \omega_i \times \left[ \frac{F_{\mathrm{obs}} (\lambda_i) - F_{\star} (\lambda_i) - F_{\mathrm{model}} (\lambda_i)}{\sigma_i} \right] ^2,
\end{equation}
where $F_{\mathrm{obs}}$ is the observed flux with its associated uncertainty $\sigma_i$, $F_{\star}$ the stellar contribution, and $F_{\mathrm{model}}$ the modeled thermal emission of the debris disk. The $\omega_i$ values are weights to the observed data points at wavelength $i$. This weighting is designed to account for the fact that we simultaneously model broadband photometric observations (e.g., {\it Herschel}/PACS) and spectroscopic observations ({\it Spitzer}/IRS): two adjacent points in the IRS spectrum do not have the same significance as PACS 70 and 100\,$\umu$m observations. We therefore opt for a similar strategy to the one described in \citet{Ballering2013}. We compute the average spectral resolution of the IRS data and compare it to the equivalent widths of the broadband filters. One IRS point will have a weight $\omega = 1$ and broadband photometric point will have a weight equal to the number of IRS points that would fit in the corresponding equivalent width of the filter. In Table\,\ref{tab:flux}, we report the equivalent widths for the various instruments used to model the thermal emission in the far-IR. These values were taken from the Spanish Virtual Observatory filter profile service\footnote{http://svo2.cab.inta-csic.es/theory/fps3/}. For the ALMA Band\,6 observations, we assumed the equivalent width to be equal to the spectral range ($105$\,$\umu$m).

To account for different mineralogical components, we use amorphous silicate grains of olivine stoichiometry \citep[MgFeSiO$_4$, ][$\rho = 3.5$\,g.cm$^{-3}$]{Dorschner1995} and amorphous water ices \citep[][$\rho = 1.2$\,g.cm$^{-3}$]{Li1998}. To mimic porosity, we also add the fraction of vacuum to the pool of free parameters. We mix the optical constants of the different dust components, using the Bruggeman mixing theory, and the Mie theory to compute the absorption coefficients $Q_{\mathrm{abs}}$. Since computing $Q_{\mathrm{abs}}$ for large grain sizes can be the bottle-neck when computing thousands of models, we first created a library of opacity files ($s_{\mathrm{min}} = 0.01\,\umu$m and $s_{\mathrm{max}} = 5$\,mm, with steps of 10\% for the water ices, and porosity fractions). During the modeling process, the appropriate $Q_{\mathrm{abs}}$ are drawn and interpolated for the proper grain size distribution and composition.

The free parameters include the grain size distribution ($s_{\mathrm{min}}$ and $p$, we fix $s_{\mathrm{max}} = 5$\,mm), the relative fractions for the fractions of amorphous water ices and porosity, and the inner and outer slopes for the dust density ($\alpha_{\mathrm{in}}$ and $\alpha_{\mathrm{out}}$, respectively). We fix the reference radius $r_0$ to the best-fit results of the modeling of the SPHERE observations. For each set of parameters the dust mass M$_{\mathrm{dust}}$ is found by scaling the thermal emission of the model to the observed SED (similarly to Eq.\,\ref{eqn:minimize}). The values for the dust mass are saved for posterior estimation of their corresponding uncertainties. To speed up the modeling of the SED, we neglect the non-azimuthal dust density distribution derived from the analysis of the SPHERE data and assume the disk to be a circular ring. We use the \texttt{emcee} package to search for the most probable model, using 100 walkers, burning the first 500 runs for each walker, and then running the chain for 1\,000 more iterations. The acceptance fraction at the end of the run is of $0.28$ (and a maximum auto-correlation time across all parameters of 72 steps).

\subsection{Results}

\begin{figure*}[tbh]
\includegraphics[width=2\columnwidth]{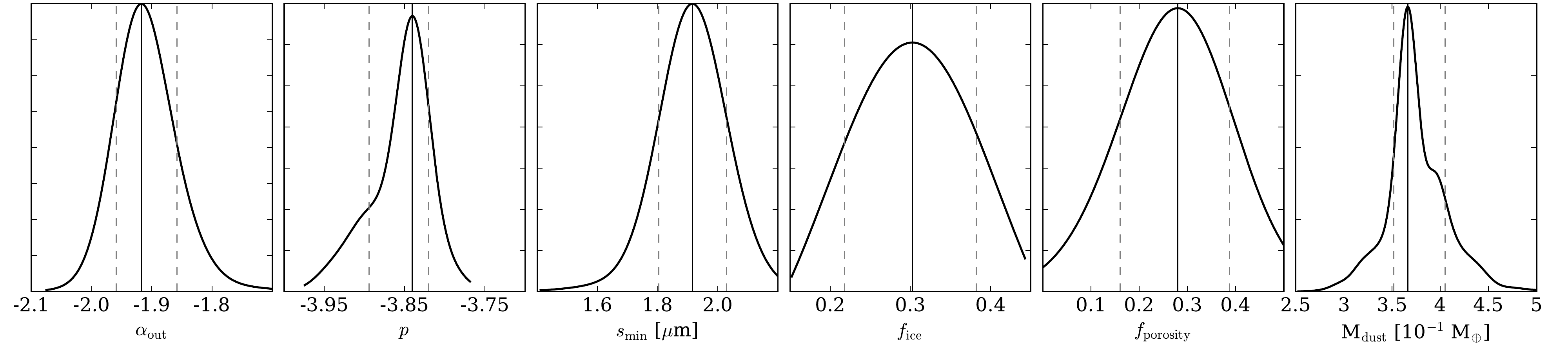}
\caption{Posterior probability distributions for the modeling of the SED. The solid lines indicate the most probable value, and the vertical dashed lines indicate the derived uncertainties.}
\label{fig:pdf_SED}
\end{figure*}

\begin{figure}
\includegraphics[width=1.0\columnwidth]{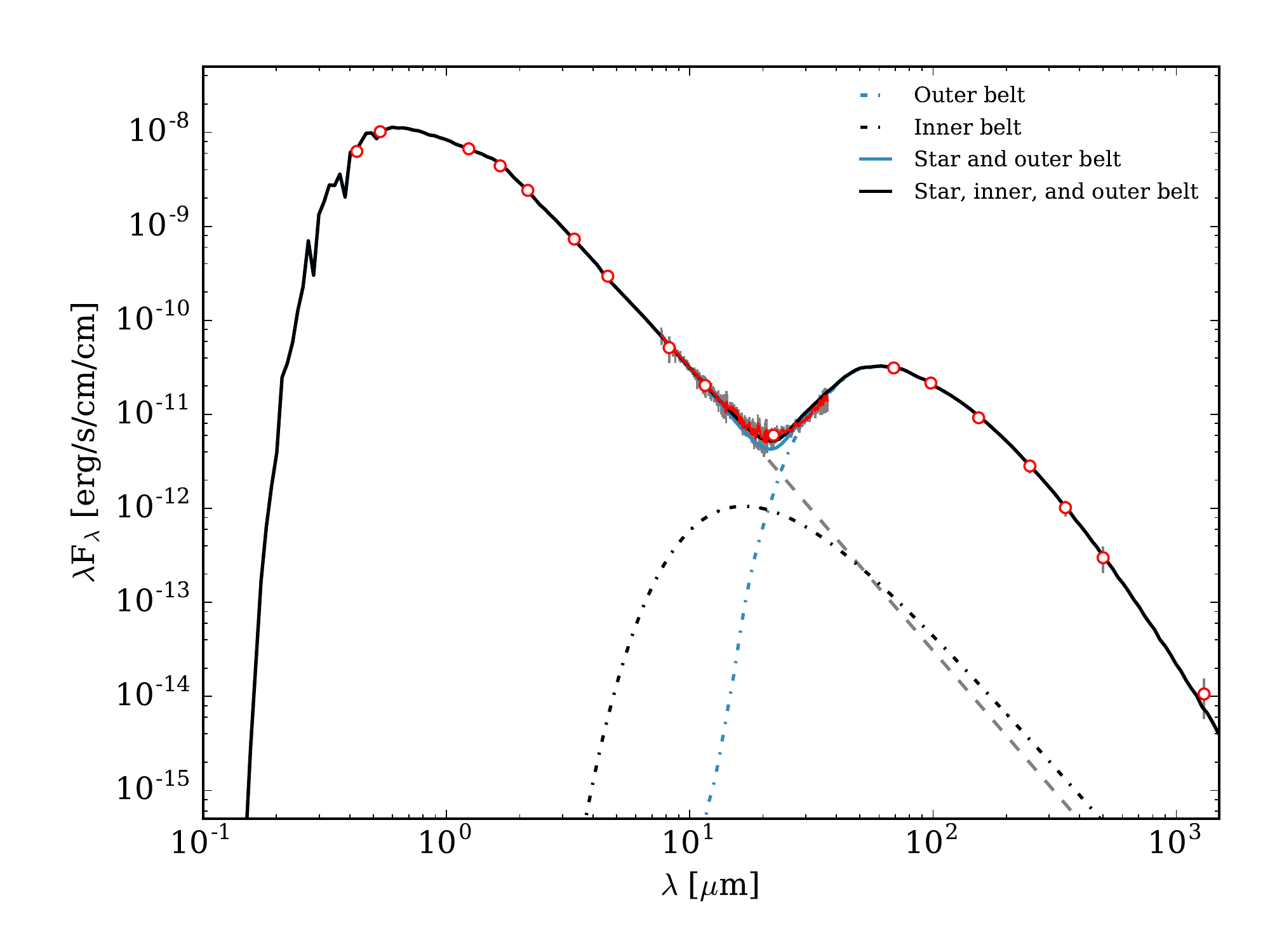}
\caption{Spectral energy distribution of HD\,61005. Photometric measurements are shown in red, along with the {\it Spitzer}/IRS spectrum (uncertainties, shown in gray, are $3\sigma$, most of them being smaller than the symbol). The dashed gray line is the photospheric model, the black and cyan lines are the best fit models (including the stellar contribution, but with and without the inner component, see text for details).}
\label{fig:SED}
\end{figure}

\begin{table}
\centering
\caption{Best-fit results for the modeling of the SED.\label{tab:res_SED}}
\begin{tabular}{@{}lccc@{}}
\hline\hline
Parameter & Uniform prior & $\sigma_{\mathrm{kde}}$ & Best-fit value \\
\hline
$\alpha_{\mathrm{out}}$                      & $[-10, -0.5]$  & $0.04$ & $-1.92_{-0.04}^{+0.06}$ \\
$p$                                          & $[-5.0, -2.0]$ & $0.02$ & $-3.84_{-0.05}^{+0.02}$ \\
$s_{\mathrm{min}}$ [$\umu$m]                 & $[0.01, 10]$   & $0.1$  & $1.9_{-0.1}^{+0.1}$ \\
$f_{\mathrm{ice}}$                           & $[0, 0.95]$    & $0.1$  & $0.3_{-0.1}^{+0.1}$ \\
$f_{\mathrm{porosity}}$                      & $[0, 0.95]$    & $0.1$  & $0.3_{-0.1}^{+0.1}$ \\
M$_{\mathrm{dust}}$ [10$^{-1}$ M${_\oplus}$] & -              & $0.1$  & $3.7_{-0.1}^{+0.4}$ \\
\hline
\end{tabular}
\end{table}

Table\,\ref{tab:res_SED} summarizes the best fit results and the derived uncertainties and projected posterior probability distributions for each parameter are displayed in Fig.\,\ref{fig:pdf_SED}. The SED of HD\,61005, along with the photometric and spectroscopic observations, the photospheric model, and the best-fit model are shown in Fig.\,\ref{fig:SED}. The shape of the mid- to far-IR excess is well reproduced by our model, but the turn-off point, where the excess emission starts, is not well matched (near $\lambda \sim 20$\,$\umu$m). This hints towards the presence of an additional component, an inner disk that has also been postulated by many authors in the literature (e.g., \citealp{Morales2011,Ballering2013,Ricarte2013,Chen2014}). Even when using detailed dust properties (instead of Planck functions at single temperatures) to model the excess, the best-fit model still underestimates the flux in the mid-IR beyond $3\sigma$. To constrain the properties of an additional belt, we fit a Planck function to the residuals (the only free parameter being the temperature, the scaling to the residuals being done similarly to Eq.\,\ref{eqn:minimize}). We find the temperature of the Planck function to best reproduce the residuals to be $\sim 220$\,K. In Fig.\,\ref{fig:SED}, the Planck function is shown as a dotted-dashed black line and the final best-fit model (stellar contribution plus the inner and outer belts) as the solid black line. To assess the relevance of adding the inner disk we follow a similar approach as in \citet[][and references therein]{Moor2015} and use the Akaike Information Criterion (AIC). We find that the addition of the inner component significantly improves the final model (the AIC including the inner belt is about $80$\% of the AIC without it), and therefore is deemed necessary to reproduce the whole SED of the disk around HD\,61005.

Our modeling results suggest that the outer slope of the dust density distribution is relatively shallow, which would indicate that most of the grains responsible for the emission are located in an extended disk. The minimum grain size is found to be $\sim 1.9$\,$\umu$m, and the grain size distribution has a steep slope of $p \sim -3.84$. The composition of the grains would be a mixture of amorphous silicate grains with small fraction of water ices and porosity. The total dust mass (between $s_{\mathrm{min}}$ and $s_{\mathrm{max}}$) is of the order of $0.37$\,M$_{\oplus}$.

\section{Discussion}\label{sec:discussion}

In the following, we try to put together the results of the modeling of the ALMA, SPHERE datasets as well as of the SED. We discuss what can be inferred regarding the dust properties, the gas content, the morphology of the disk to better characterize what is happening in this system. Because we modeled several datasets (ALMA, SPHERE, SED), Table\,\ref{tab:summary} summarizes which parameters are constrained with which datasets, to clarify the discussion.

\begin{table*}
\centering
\caption{Summary of the modeling strategy adopted in this study. Even though the SPHERE DPI and ADI data are fitted simultaneously, we separated the entries to better see dependencies. A dash denotes that the parameter is irrelevant during the fitting of a given dataset.\label{tab:summary}}
\begin{tabular}{@{}lccccc@{}}
\hline\hline
Parameter & ALMA data & Polarized intensity & SPHERE DPI & SPHERE ADI & SED  \\
          & [1]       & [2]                 & [3]        & [4]        & [5]  \\
\hline
$r_0$ [au]                          & Fitted         & -            & Fitted       & Fitted & Fixed to [3,4] \\
$\alpha_{\mathrm{in}}$              & $5$            & -            & $5$          & $5$    & $5$ \\
$\alpha_{\mathrm{out}}$             & Fitted         & -            & Fitted       & Fitted & Fitted \\
$\phi$ [$^{\circ}$]                 & Fitted         & Fixed to [1] & Fitted       & Fitted & - \\
$i$ [$^{\circ}$]                    & Fitted         & Fixed to [1] & Fitted       & Fitted & - \\
$f_{\mathrm{1300}}$ [mJy]           & Fitted         & -            & -            & -      & - \\
$s_{\mathrm{min}}$                  & $0.1$\,$\umu$m & Fitted       & Fixed to [2] & -      & Fitted \\
$s_{\mathrm{max}}$                  & $5$\,mm        & Fitted       & Fixed to [2] & -      & $5$\,mm \\
$f_{\mathrm{porosity}}$             & $0$            & Fitted       & Fixed to [2] & -      & Fitted \\
$e$                                 & $0$            & -            & Fitted       & Fitted & - \\
$\Phi_{\mathrm{e}}$ [$^{\circ}$]    & -              & -            & Fitted       & Fitted & - \\
$\eta$                              & $1$            & -            & Fitted       & Fitted & - \\
$\Phi_{\mathrm{\eta}}$ [$^{\circ}$] & -              & -            & Fitted       & Fitted & - \\
$w$ [$^{\circ}$]                    & -              & -            & Fitted       & Fitted & - \\
$\psi$                              & $0.05$         & -            & Fitted       & Fitted & - \\
$g$                                 & -              & -            & -            & Fitted & - \\
$f_{\mathrm{ADI}}$                  & -              & -            & -            & Fitted & - \\
$p$                                 & $-3.5$         & $-3.5$       & $-3.5$       & $-3.5$ & Fitted \\
$f_{\mathrm{ice}}$                  & $0$            & $0$          & -            & -      & Fitted \\
M$_{\mathrm{dust}}$ [M$_\oplus$]    & -              & -            & -            & -      & Fitted \\
\hline
\end{tabular}
\end{table*}

\subsection{Dust properties}\label{discuss:dust}

\subsubsection{Self-subtraction corrected phase function}

\begin{figure}
\includegraphics[width=\columnwidth]{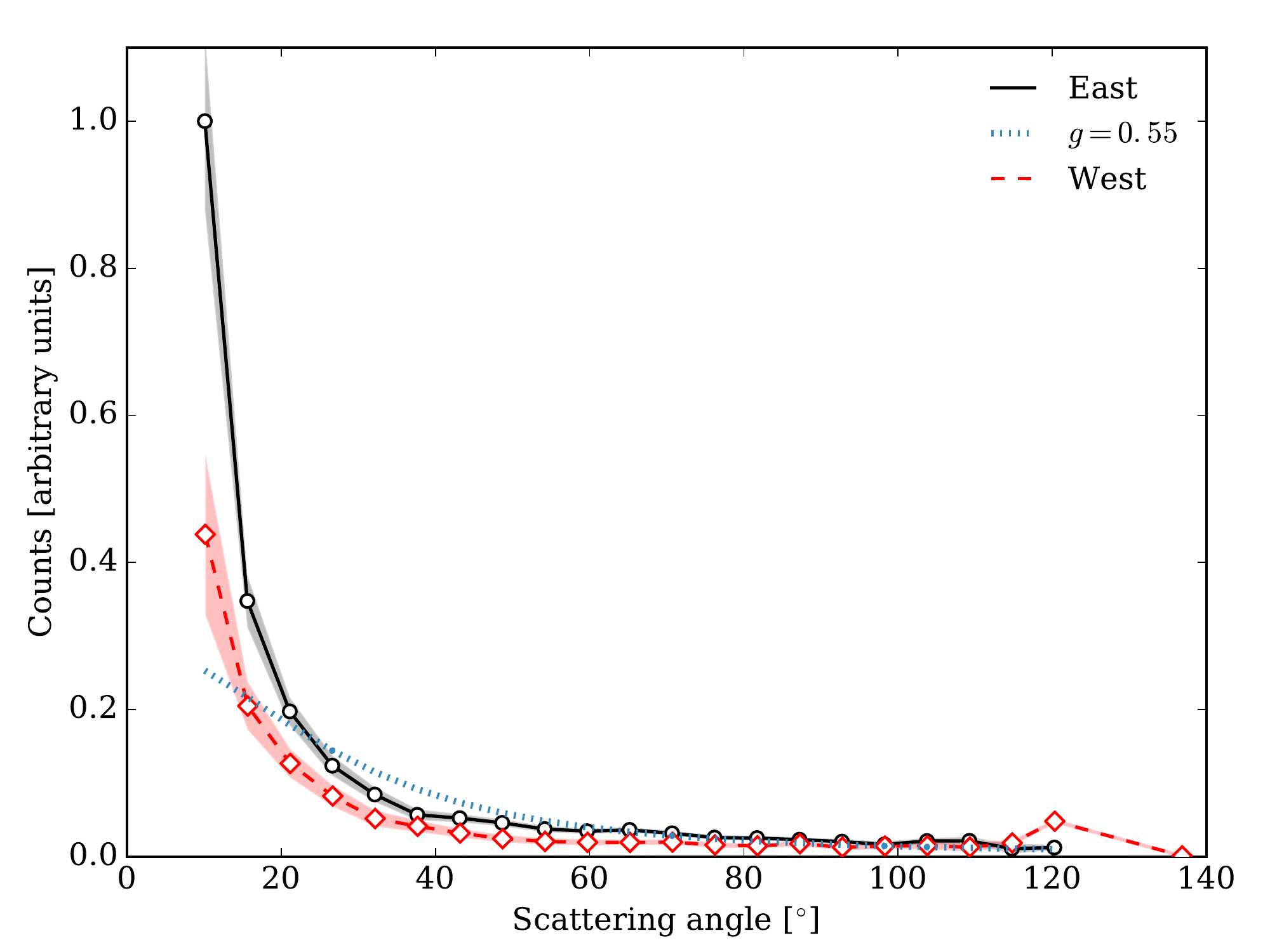}
\caption{Phase function derived from the ADI $H$-band observations, corrected for self-subtraction effects (see text for details).}
\label{fig:pfunc_ADI}
\end{figure}

With a satisfying model for the scattered light, we can further investigate the phase function in the SPHERE ADI $H$-band observations. Indeed, the model enables us to estimate the self-subtraction introduced by the PCA analysis. To do so, we first perform the PCA on the raw data cube and save the resulting coefficients. We then take the best-fit model, produce a data cube at the proper parallactic angles, project it into the new orthogonal basis found for the observations, and multiply this projection with the same coefficients as for the raw observations (e.g., \citealp{Soummer2012}). This provides us with an estimate of the signal from the disk that is subtracted when doing the PCA. The attenuation is estimated by the ratio $r_{\mathrm{corr}}$ between this estimate and the original model. When deriving the phase function for the ADI dataset (similarly to Section\,\ref{sec:pfunc}), each pixel is multiplied by the quantity $1/(1 - r_{\mathrm{corr}})$. Figure\,\ref{fig:pfunc_ADI} shows the phase function derived for the east and west sides (black and red, respectively). The bump at $\sim 120^{\circ}$ on the west side is an artifact introduced by $r_{\mathrm{corr}}$ (low signal in the disk model leads to an erroneous correction factor). In Fig.\,\ref{fig:pfunc_ADI}, we also show the phase function derived during the modeling (for $g = 0.55$), scaled to the phase function of the east for scattering angles larger than $40^{\circ}$. The inferred HG function appears to be a relatively good representation of the corrected phase function for scattering angles larger than $\sim 50^{\circ}$, but severely fails to reproduce it at smaller angles. This would be an explanation for the strong residuals along the semi-minor axis in Fig.\ref{fig:residuals_H}. Overall, this suggests quite extreme forward scattering in the disk around HD\,61005 - a result that remains model-dependent since the PCA attenuation is estimated from our results (see also Milli et al. submitted for a similar discussion for the disk around HR\,4796\,A). Nonetheless, the strong peak of scattering for small angles (down to $\sim 10^{\circ}$) strongly points towards the presence of large dust grains in the debris disk. However, this seems in contradiction with the prior analysis of the DPI dataset where we had to actually reject the presence of large dust grains ($s \sim 0.3$\,$\umu$m). For the SED, we need a whole range of sizes (from $\umu$m- to mm-sized grains), but the emission is actually dominated by the smaller grains (discussed further in this section). Numerous studies of debris disks around different stars (e.g., \citealp{Rodigas2015}, \citealp{Lebreton2012}, \citealp{Milli2015}; Milli et al. submitted) already reported and discussed similar issues in reconciling the modeling results of different kind of observations. Throughout this paper, for their computational merits, we have either used the Mie theory or the HG prescription while both of them may not be accurate descriptions of the nature of the dust grains. 

\subsubsection{Reconciling different observations}

To try to reconcile the different dust properties inferred from our modeling results, the radial extent of the disk is interesting to discuss. The ALMA observations point towards a narrow belt ($\alpha_{\mathrm{out}} \leq -5$), while the SED, ADI, and DPI observations suggest that the disk is extended ($\alpha_{\mathrm{out}} \sim -2$ or $-3$). To assess the typical grain sizes we probe with the SED, we computed the integrated infrared luminosity for each grain size bins between $s_{\mathrm{min}}$ and $s_{\mathrm{max}}$, and built the cumulative distribution function. We find that more than $98$\% of the disk infrared luminosity is accounted for by dust grains smaller than $5$\,$\umu$m. The rest of the (very steep) grain size distribution contributes marginally to the total emission. This means that for the DPI observations and the SED, we are observing an extended disk containing small dust grains. We computed the $\beta$ ratio between the radiation pressure and gravitational forces as in \citet{Burns1979}, assuming $L_* = 0.58$\, $L_{\odot}$ and $M_* = 1.1$\,$M_{\odot}$. For the dust composition found from the SED modeling, we find that grains larger than $s_{\mathrm{blow}} \sim 0.8$\,$\umu$m, for which $\beta \leq 0.5$,  should remain on bound orbits (assuming they were released on circular orbits). There is a small discrepancy between $s_{\mathrm{blow}}$ and the minimum grain size we found ($s_{\mathrm{min}} \sim 2$\,$\umu$m) but as discussed in \citet{Pawellek2014}, this could be caused by the assumption of spherical grains when modeling the SED. Assuming our inferred dust composition is not too far off compared to the real composition of the dust grains, it seems that radiation pressure should be efficient for (sub-)\,$\umu$m-sized grains. Consequently, the parent belt (inferred from the ALMA data) would be fairly narrow, but small grains placed on eccentric orbits (or even unbound) owing to radiation pressure make the disk look more extended at near-IR wavelengths. 

The strong forward scattering peak derived from the corrected phase function from the ADI observations suggests the presence of large grains in an extended disk, which seems in contradiction with the aforementioned scenario. Nonetheless, according to \citet{Min2016}, large dust aggregates behave like large grains (the size of the whole aggregate) in scattered light and like small grains (the size of the individual monomers) for polarized light. Such aggregates also display mild backward scattering that we do not detect with our observations but may be the consequence of low S/N along the back side of the disk.

Therefore, we hypothesize that the parent planetesimals are arranged in a narrow belt (traced by ALMA) and that the small-size end of the grain size distribution consists of dust aggregates that are pushed away by radiation pressure. This makes the disk appear extended in the SPHERE observations and when modeling the SED. For future studies of this system, it would be interesting to implement a radial segregation of the grain size distribution (varying d$n(s)$ as a function of $r$, see for instance \citealp{Stark2014} for a discussion of the phase function depending on the radial distance). 

\subsection{Gas mass upper limits}\label{sec:CO}

No CO emission was detected in the ALMA data, and we follow a similar procedure to that of \citet{Matra2015} and \citet{Moor2015b} to derive upper limits for the CO gas mass. We extract the spectrum from the datacube reduced within CASA, centered around $230.5$\,GHz ($^{12}$CO 2-1). The upper limit for the CO mass was determined as
\begin{equation}
M_{\mathrm{CO}} = \frac{4 \pi m d_{\star}^2}{h \nu_{2-1} A_{2-1}}\frac{S_{2-1}}{x_{2-1}},
\end{equation}
where $m$ is the mass of the CO molecule, $d_{\star}$ the distance of the star, $\nu_{2-1}$ the frequency of the transition, $A_{2-1}$ the Einstein coefficient\footnote{Taken from the SPLATALOGUE catalog at \url{http://www.cv.nrao.edu/php/splat/}}, $x_{2-1}$ the fractional population of the upper level, and $S_{2-1}$ the observed integrated line flux. The fractional level was calculated assuming local thermal equilibrium, using the Boltzmann equation, with a temperature of $20$\,K (as discussed in \citealp{Moor2015b}, $M_{\mathrm{CO}}$ is not strongly dependent on the temperature). $S_{2-1}$ was computed from the spectrum as $S_{\mathrm{rms}}\Delta v \sqrt N$, where $\Delta v$ is the channel velocity width, $N$ the number of channels over a width of $15$\,km.s$^{-1}$ (centered at the local standard of rest velocity, $v_{\mathrm{LSR}}$), and $S_{\mathrm{rms}}$ the standard deviation of the spectrum within that velocity range. From the heliocentric radial velocity of $22.5$\,km.s$^{-1}$ (\citealp{Desidera2011}), we obtain a $v_{\mathrm{LSR}}$ of $3.68$\,km.s$^{-1}$. Assuming optically thin gas, this leads to an upper limit of $6.9\times 10^{-7}$\,M$_{\oplus}$. As discussed in \citet{Matra2015}, the LTE hypothesis may not hold in low-density environment and our upper limit could underestimate the CO gas mass. In the ISM the CO/H$_2$ abundance is $10^{-4}$ and with our inferred dust mass of $\sim 0.37$\,M$_{\oplus}$ (for our grain size distribution), this would lead to a gas-to-dust ratio that is much smaller than unity. However, it is unlikely that this ratio will remain the same in a circumstellar disk because of effects such as photo-dissociation. Overall, with the available observations leading to a non-detection, and given the several assumptions made (LTE, CO/H$_2$ ratio) we can hardly conclude on the gas-to-dust ratio on the debris disk. This question should be addressed by future, deeper CO observations (or other species than the CO molecule).

\subsection{Stirring of the planetesimals}\label{sec:stirring}

In the last few years, theoretical works have aimed at characterizing the time evolution of debris disks (e.g., \citealp{Kenyon2006,Kenyon2008}). One of the purpose of these studies has been to better understand the growth of planetesimals to sizes of about a thousand km, in the framework of terrestrial planetary formation. It is believed that once these large bodies have been formed, they stir the population of smaller planetesimals. The stirring increases the relative velocities of these planetesimals and increases their chances of collisions, therefore initiating a collisional cascade. According to these models, collisions between planetesimals become destructive only once a Pluto-sized object is formed. Since the timescale for the growth of planetesimals scales with the distance $r$ to the star, self-stirring is thought to be an inside-out process. Here, we follow a similar approach to that described in \citet{Moor2015}, who studied several debris disks spatially resolved with the {\it Herschel} observatory. They examined if the disks in their sample are consistent with a self-stirring scenario. To achieve this, they compared the timescale for forming a 1\,000\,km-sized object in a self-stirred debris disk with the age of the stars. To quantify this timescale $t_{\mathrm{1000}}$ (in Myr), we used Eq.\,41 from \citet{Kenyon2008}, which we recall below:
\begin{equation}
  t_{\mathrm{1000}} = 145x_{\mathrm{m}}^{-1.15} (r/80)^3 (2M_{\odot} / M_{\star})^{3/2}\,\mathrm{[Myr]},
\end{equation}
where $r$ is the distance of the dust belt, and $x_{\mathrm{m}}$ is a scaling factor to parametrize the disk's initial surface density (the higher, the more massive). \citet{Mustill2009} argue that $x_{\mathrm{m}}$ values larger than 10 are highly unlikely since the initial disk would have been gravitationally unstable. Assuming that the timescale for the growth of the planetesimals is the age of the system (i.e., $t_{\mathrm{1000}} = 40$\,Myr), we find that $x_{\mathrm{m}}$ should be of the order of about 3.3 (for $r = 61$\,au). This value would suggest that either the primordial disk was $\sim$\,3 times more massive than the minimum-mass solar nebula, or that other sources of stirring (e.g., induced by a planet) need to be taken into consideration for this system. Debris disks with known planets, such as the ones studied in \citet[][e.g., HR\,8799, $\beta$\,Pictoris, HD\,95086]{Moor2015}, have much larger $x_{\mathrm{m}}$ values ($\geq 10$), while most of the debris disks usually have values for $x_{\mathrm{m}}$ smaller than $3$. It is therefore reasonable to assume that the disk around HD\,61005 is bright because self-stirring by planetesimals has recently started. It does not imply that alternative stirring mechanisms should not be considered, but based on the distance of the dust belt and the age of the system, it is not mandatory to invoke the presence of a massive planet to explain what is currently observed.

\subsection{An eccentric and asymmetric disk}

\begin{figure*}
\includegraphics[width=2\columnwidth]{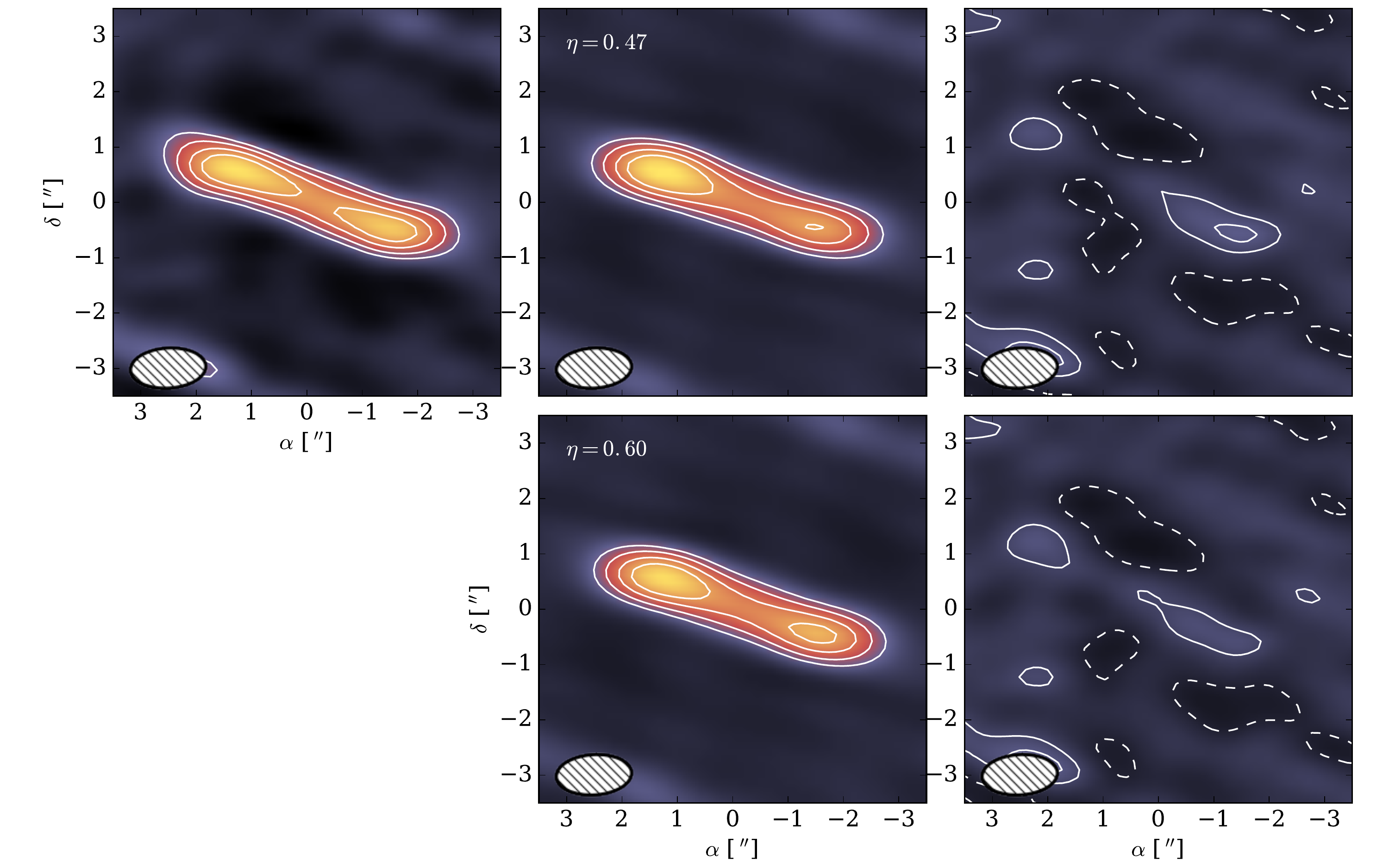}
\caption{From left to right: observations, models including some of the results of the SPHERE modeling, and residuals. Top row is for $\eta = 0.47$ and the bottom row for $\eta = 0.6$. For all panels the color map is a linear stretch between $-0.27$ and $1.23$\,mJy/beam. For both the observations and the model the contours are set at [$3$, $5$, $7.5$, $10$]$\sigma$, and [$-2$, $-1$, $1$, $2$]$\sigma$ for the residuals. The beam size is shown in the lower left corner of each panels.}
\label{fig:damping_ALMA}
\end{figure*}

With the new SPHERE observations, we confirm the findings of \citet{Buenzli2010} that the disk displays a strong brightness asymmetry. Both the ADI and DPI observations show that the eastern side is brighter than the western side at an unprecedented angular resolution. The asymmetry is detected both in scattered light along the semi-minor axis (ADI dataset) and in scattered and polarized light along the semi-major axis (NACO observations of \citealp{Buenzli2010} and DPI observations, respectively). When modeling the observations, we had to include a density damping factor along the azimuthal direction to reproduce the SPHERE observations. From preliminary tests, the DPI dataset (in which the brightness asymmetry is most visible) cannot be properly reproduced without these azimuthal variations. Therefore, we argue it can hardly be a scattering artifact and that it is, indeed, most likely related to a dust density enhancement in the disk.

On the other hand, the ALMA observations do not display such a strong asymmetry at first glance. To further investigate if the best fit model to the SPHERE observations is compatible with the ALMA data, we generated models including eccentricity and azimuthal variations. We took the best fit model to the ALMA data (Table\,\ref{tab:alma}), and the following additional parameters from Table\,\ref{tab:res_H}: $e = 0.093$, $\Phi_{\mathrm{e}} = 127^{\circ}$, $w = 52^{\circ}$, $\Phi_{\mathrm{\eta}} = 138^{\circ}$, and two values for $\eta$ ($0.47$ and $0.6$). Figure\,\ref{fig:damping_ALMA} shows the comparison between these models and the observations. For $\eta = 0.47$ (derived from the SPHERE observations), the brightness of the western side is slightly underestimated but the residuals barely reach the $2\sigma$ level. For $\eta = 0.6$, the model becomes more similar to the observations and the residuals do not reach the $2\sigma$ level. Consequently, we cannot exclude that the disk shows departure from centro-symmetry in the ALMA data, but the sensitivity of the observations does not allow us to put meaningful constraints on the azimuthal variations for the population of large grains. We can exclude damping factors of $\eta \sim 0.5$ to the $2\sigma$ level, but a value of $0.6$ cannot be ruled out based on the available observations. 

\subsubsection{Possible planet-disk interactions ?}

Massive planets are often invoked to explain the eccentricity of a debris disk since the planet may shepherd it. The case of Fomalhaut is a good example of these types of studies. The disk shows an eccentricity comparable to the one of HD\,61005 ($e \sim 0.11\pm0.1$, \citealp{Kalas2005}) and a candidate companion was detected by \citet{Kalas2008}. Studies such as the one of \citet{Beust2014} investigate in detail the possible interactions between the planet and the planetesimal in the disk. In the last two decades, many numerical works have been led to investigate how planets can shape debris disks (e.g., \citealp{Wyatt1999,Wyatt2006,Nesvold2013,Pearce2014,Faramaz2014}), or how the planetesimals evolve within their mutual gravitational interactions (e.g., \citealp{Kral2015,Jackson2014}).

Via secular interactions, an eccentric planet can shape the debris disk. Because of differential precession, the planet-disk interactions can result in a (short-lived) spiral arm (\citealp{Mouillet1997,Wyatt2005}). Then, when taking into account the effect of collisions in the disk, the spiral quickly dissipates (\citealp{Nesvold2013}), and the disk becomes eccentric (and apse-aligned with the planet's orbit). Because of the eccentricity, the disk is expected to be brighter at the pericenter (the so-called pericenter glow effect). Nonetheless, this may not be what is happening in the disk around HD\,61005. First, we did not detect any point sources in neither the $H$- nor the $K$-band ADI datasets, even with their large parallactic rotations. The exact detection limits on the masses of possible planets will be published in the survey paper of the SPHERE consortium, but preliminary tests using the \texttt{PynPoint} package (\citealp{Amara2012}) shows that we should have been able to detect planets with a contrast of about $\sim$\,10\,mag in the near-IR. At an age of about 40\,Myr, we can therefore rule out any planets with a mass larger than $\sim$\,7 Jupiter mass at a distance of $0.2^{\prime\prime}$. Second, as already discussed in \citet[][and confirmed in this study]{Buenzli2010}, the eccentricity of the disk is not sufficient for the pericenter glow to solely explain the observed brightness asymmetry. Our modeling results suggest that there is a $\sim$\,50\% reduction in density between the pericenter and the apocenter, with a peak density closer to the observer ($\Phi_{\mathrm{e}} \sim \Phi_{\eta} \sim 130^{\circ}$). Given that both $\Phi_{\mathrm{e}}$ and $\Phi_{\eta}$ are free parameters and that their most probable values are so close to each other, it strongly points towards an increase of the dust density distribution at the pericenter of the eccentric disk. Interestingly, the simulations of \citet{Pearce2014} show the opposite effect: the density is lower at the pericenter when including an eccentric planet in the disk (the particles spend more time at the apocenter than at the pericenter). 

\citet{Wyatt2006} describes the interactions between a massive planet with the dust belt, in different configurations. Whether the planetesimals are in 2:1 or 3:2 resonances with the planet, we can expect the disk to display asymmetries not only in scattered light but also at millimeter wavelengths. The large grains traced by mm observations are expected to closely follow the planetesimals that are in resonance with the planet. As assessed previously, with the current sensitivity of the mm observations, we cannot confirm nor rule out that large dust grains display departure from centro-symmetry.

Therefore, if a planet were to be detected (at other wavelengths or with other facilities), it would be interesting to confront our observations and modeling results with the orbital parameters of the planet. But in the following, we hypothesize that no massive planet was formed around HD\,61005 and we try to explain the inferred morphology of the disk under this \textit{assumption}.

\subsubsection{The impact of gas in a debris disk}

\citet{Lyra2013} show that narrow, eccentric debris disks can be the consequence of the presence of gas. While debris disks are generally thought of as depleted of gas, several recent studies have demonstrated it is not necessarily the case, even at ages as old as a few tens of Myr (e.g., \citealp{Hughes2008,Dent2014,Kospal2013,Moor2015b}). \citet{Lyra2013} suggest that the dust grains can heat up the surrounding gas via photoelectric heating. This will locally increase the gas pressure and, since the grains follow the pressure gradient, the pressure maximum will concentrate them even further. In the end, the disk may appear as narrow and eccentric, even though no planets are shaping it (although the original disk must be narrow). In Section\,\ref{sec:CO}, we concluded that the gas-to-dust ratio remains highly uncertain because of several unconstrained parameters (grain size distribution, CO/H$_2$ abundance ratio, LTE assumption). Overall, this scenario may be a viable solution to explain our findings, but it cannot be verified at the moment (the azimuthal density variation would have to be addressed as well).

\subsubsection{The aftermath of a collision between Pluto-sized bodies ?}\label{sec:collision}

An alternative possibility would be that we are witnessing an impact between two massive bodies in the debris disk. This solution is mostly interesting because we find a peak of density at the pericenter of a slightly eccentric disk. In Section\,\ref{sec:stirring}, we concluded that self-stirring by massive planetesimals may be taking place in the 40\,Myr old debris disk. At this age, and a radius of 60\,au, a few massive oligarchs may have been formed, which are disrupting the orbits of other planetesimals. The collisional cross-sections of these bodies would have increased along with the probabilities of intercepting each others' orbits. Numerical studies of these types of impact (e.g., \citealp{Jackson2014} and \citealp{Kral2015}) suggest that our results may be compatible with the scenario of an impact between two massive bodies. This type of collision would release significant amount of debris, and all of it would have to pass through the same location where the collision initially occurred. This would enhance the density and therefore the chances of subsequent collisions at the location of this pinch point. In which case, the scattering cross-section will locally increase along the eastern side compared to the western side. Similar scenarios have been postulated for the highly asymmetric disk around HD\,15115 (\citealp{Mazoyer2014}), $\beta$\,Pictoris (\citealp{Dent2014}), and HD\,181327 (\citealp{Stark2014}). According to the work of \citet{Jackson2014}, the debris disk around HD\,61005 could maintain this asymmetry for about $\sim 0.45$\,Myr at 60\,au (1\,000 orbital periods). However, in these numerical simulations, the eccentricity and overall complexity of the resulting disk highly depends on the initial conditions of the impact. Furthermore, \citet{Leinhardt2012} find that the grain size distribution of the fragments released by the collision of gravity-dominated bodies is best described by a power-law slope in $-3.85$. Our modeling results of the SED are in excellent agreement with this value, supporting the scenario that the disk mainly consists of fragments from a catastrophic collision.

Nonetheless, a challenging aspect to this postulate is the dust mass inferred from the SED modeling. Simulations of the stochastic phase of planetesimal growth by \citet{Stewart2012} suggest that collision of massive bodies should release about $\sim 15$\% of the total mass in debris. Assuming ``debris'' corresponds to grains with sizes between $2$\,$\umu$m and $5$\,mm (used to model the SED), this would correspond to a total mass of about $2.5$\,M$_{\oplus}$ for the catastrophic impact, which would make this type of event an outliner in most theoretical works about the formation of planetesimals. For a 40\,Myr old solar-mass star, the models of \citet{Kenyon2008,Kenyon2010} predict a dust mass production of the order of $\sim 0.05$\,M$_\oplus$ (for $1$\,$\umu$m\,$\leq s \leq 1$\,mm) for a disk that is initially $3$ times more massive than the MMSN (to be compared to $0.25$\,M$_\oplus$ interpolating our results to the same grain size distribution). If the primordial disk around HD\,61005 was indeed several times more massive than the MMSN (see Section\,\ref{sec:stirring}), it would evolve more rapidly and massive bodies would have had the time to form. Overall, if this interpretation of the multi-wavelength observations of the Moth holds, it would suggest that massive bodies can still be formed at a large distance from the star at an age of $\sim 40$\,Myr. This would provide valuable constraints on the formation and evolution of planetesimals in the first 100\,Myr.

The presence of spatially separated belts (inner and outer belts) is often discussed in the context of planet-disk interactions (e.g., \citealp{Su2013}). Massive planets may scatter away any close-by debris, hence opening a dust-free region in the disk. Planets may also be able to scatter debris from the outer regions and send them inwards, which would result in a bi-modal distribution of the dust (e.g., \citealp{Bonsor2012}). In the case of HD\,61005, we do not have strong constraints on the location of the inner belt. But, given that the evolution of a debris disk is an inside-out process (\citealp{Kenyon2008}), it is unlikely the inner belt is primordial. Given the age of the star, the inner regions should already have been depleted of any detectable amount of small dust grains. Still, assuming that there are no massive planets around HD\,61005, is it possible to reconcile the presence of an analog of the asteroid belt with the above interpretation ? If it is indeed the case that a few oligarchs have formed in the main belt at 60\,au, the disk would be in an active and chaotic state. Some of these massive bodies may have collided in the main belt, but some may have scattered debris inwards. However, simulations, such as the ones presented in \citet{Jackson2014} and \citet{Kral2015}, do not display cleared gaps. Consequently, under our assumption that there are no giant planets, it seems that our interpretation of the dynamical evolution of the outer regions is not directly related with what is taking place in the inner regions. It may well be that telluric planets have formed in the inner regions and further investigation remains necessary to better constrain the location of the innermost belt.

\section{Conclusion}\label{sec:conclusion}

In this study, we presented new VLT/SPHERE and ALMA observations of the debris disk around the 40\,Myr-old solar-type star HD\,61005. We modeled both the radial and azimuthal distribution of the dust grains. We find that the disk is slightly eccentric ($e \sim 0.1$), with a dust density two times smaller at the apocenter compared to the pericenter. The sensitivity of the ALMA observations cannot confirm nor rule out an asymmetric disk for large mm-sized dust grains. We confirm that the eccentricity of the disk is not sufficient to explain the known brightness asymmetry in the near-IR. Thanks to the multi-wavelength analysis, we highlight a possible radial segregation of the grain size distribution since the disk appears more extended at near-IR wavelengths compared to far-IR and mm wavelengths. The parent planetesimals are confined within a narrow belt (with a semi-major axis of about $61$\,au), while small dust grains (or dust aggregates) would be pushed away by radiation pressure. No CO gas was detected with the ALMA observations and we did not detect any giant planets around the star. For this reason, we attempt to explain the morphology of the disk without invoking planet-disk interactions. We propose that we could be witnessing the aftermath of collision(s) between massive bodies in the main belt. The debris disk may be bright because self-stirring recently started to become significant in the belt, and it triggered a massive collision. The location of the collision became a pinch point where all debris must go through at each orbit, hence locally increasing the dust density and the scattering cross-section, as well as the chances of subsequent collisions. Nonetheless, this scenario would require a massive primordial disk to reconcile numerical simulations of debris disk evolution with the inferred dust mass. Finally, the presence of an inner belt would be consistent with the modeling of the SED, and we conclude its presence and dynamical evolution is unrelated to the evolution of the outer belt.

\begin{acknowledgements}
We would like to thank the anonymous referee for the valuable feedback we received. The comments improved the discussion of the paper, mostly with respect to the scenario of a massive collision. They also helped with the presentation of the modeling strategy for all the different datasets presented in this paper. This paper makes use of the following ALMA data: ADS/JAO.ALMA 2012.1.00437.S. ALMA is a partnership of ESO (representing its member states), NSF (USA) and NINS (Japan), together with NRC (Canada), NSC and ASIAA (Taiwan), and KASI (Republic of Korea), in cooperation with the Republic of Chile. The Joint ALMA Observatory is operated by ESO, AUI/NRAO and NAOJ. J. Olofsson, H. Avenhaus, C. Caceres, H. Canovas, M. R. Schreiber, A. Bayo, and S. Casassus acknowledge support from the Millennium Nucleus RC130007 (Chilean Ministry of Economy). J. Olofsson acknowledges support from ALMA/CONICYT Project 31130027, H. Avenhaus the support from FONDECYT 2015 Postdoctoral Grant 3150643, C. Caceres the support from CONICYT FONDECYT grant 3140592, H. Canovas the support from ALMA/CONICYT (grants 31100025 and 31130027), and A. Bayo the financial support from the Proyecto Fondecyt Iniciaci\'on 11140572. This work was supported by the Momentum grant of the MTA CSFK Lend\"ulet Disk Research Group. A. Mo\'or acknowledges support from the Bolyai Research Fellowship of the Hungarian Academy of Sciences. J. Olofsson would like to thank Daniel Foreman-Mackey for his help with the parallelization of the code with \texttt{emcee} and Mark Booth for valuable discussions about ALMA data. This research made use of Astropy, a community-developed core Python package for Astronomy (Astropy Collaboration, 2013). This research has made use of the SIMBAD database (operated at CDS, Strasbourg, France). The authors have used the TOPCAT software (\citealp{Taylor2005}) in this work.
\end{acknowledgements}

\bibliographystyle{aa}


\appendix
\section[]{Description of the code}\label{app:DDIT}

\subsection{Computing the spectral energy distribution}

\begin{figure}
\includegraphics[width=1.0\columnwidth]{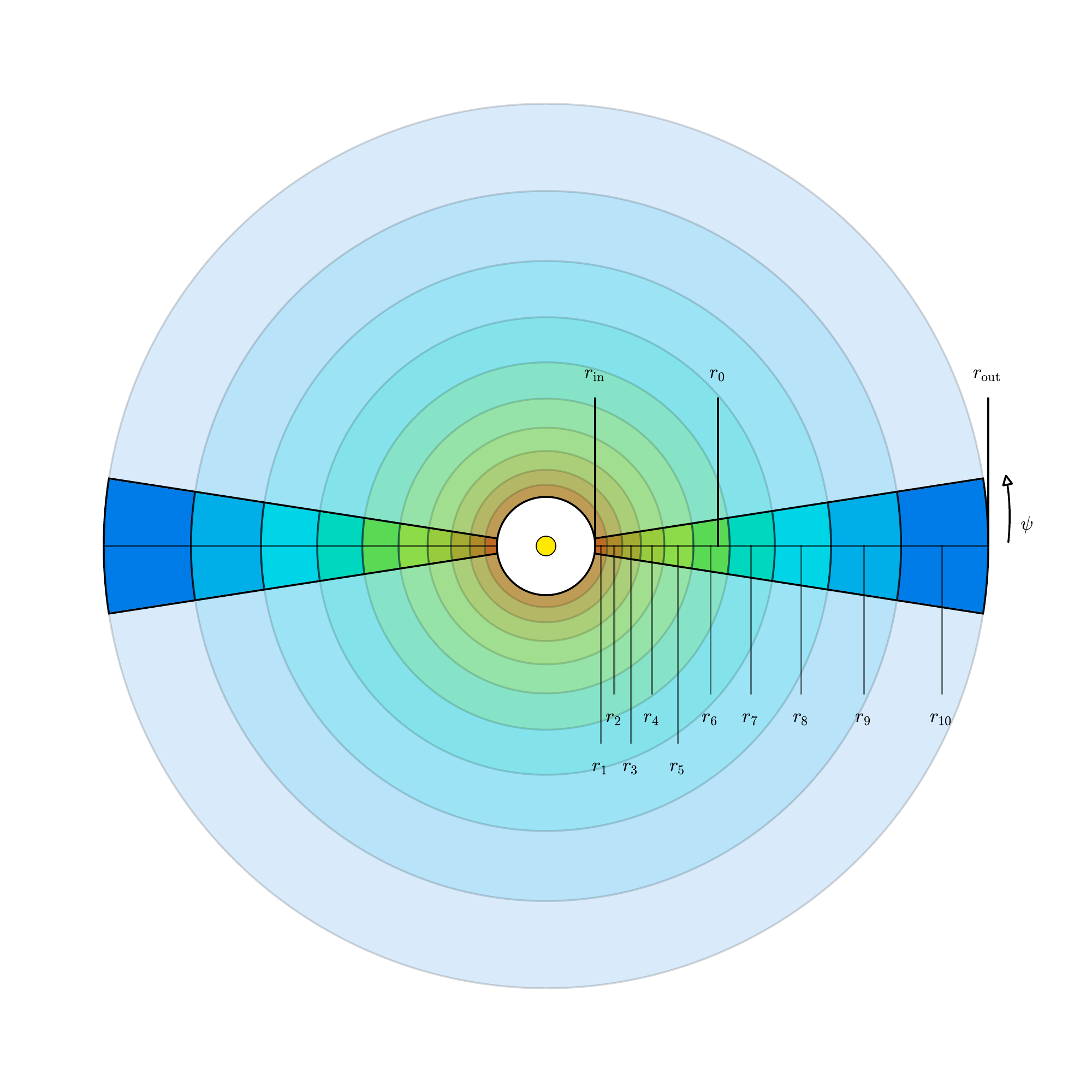}
\caption{Schematic view of how a disk, seen edge-on, is described in the DDiT code. Different shells ($n_{\mathrm{r}} = 10$) are represented in different fading colors, while dust belts are represented without transparency (see text for details).}
\label{fig:ddit}
\end{figure}

The DDiT code is mostly based on the code already described in \citet[which was based on the ``Debris disk radiative transfer simulation tool'' described in \citealt{Wolf2005}]{Olofsson2012,Olofsson2013}. However, as opposed to the work presented in \citet{Olofsson2012,Olofsson2013}, only one dust species per model is considered in this study. The dust properties are defined by the optical constants, the density ($\rho$), the minimum and maximum grain sizes ($s_{\mathrm{min}}$ and $s_{\mathrm{max}}$, respectively), and the slope $p$ of the grain size distribution (d$n(s) \propto s^p \mathrm{d}s$, $p < 0$). The absorption efficiencies, $Q_{\mathrm{abs}} (\lambda, s)$, are computed using the Mie theory for $n_{\mathrm{g}} = 100$ grain sizes logarithmically spaced between $s_{\mathrm{min}}$ and $s_{\mathrm{max}}$.

Concerning the disk model itself, this is defined by the following parameters: a reference radius $r_0$, two power-law indexes $\alpha_{\mathrm{in}} > 0$ and $\alpha_{\mathrm{out}} < 0$, the opening angle of the disk ($\psi = \mathrm{arctan}(h/r)$, where $h$ is the height from the midplane), and the total dust mass $M_{\mathrm{disk}}$. We first define the inner and outer radii of the disk by finding $r_{\mathrm{in}} = r_0 \times C^{(1/\alpha_{\mathrm{in}})}$ and $r_{\mathrm{out}} = r_0 \times C^{(1/\alpha_{\mathrm{out}})}$, where $C$ is a threshold value for the minimum density that we choose to be $C = 10^{-2}$. The code is 1D and the disk is divided into $n_{\mathrm{r}}$ spherical shells centered at $r_{\mathrm{j}}$, logarithmically spaced between $r_{\mathrm{in}}$ and $r_{\mathrm{out}}$ (see Fig.\,\ref{fig:ddit}).

The dust number density distribution for the shell at distance $r_{\mathrm{j}}$ and the grain size $s_{\mathrm{k}}$, $N_{\mathrm{dens}}(r_{\mathrm{j}},s_{\mathrm{k}})$, is defined as
\begin{equation}\label{eqn:dens}
    N_{\mathrm{dens}}(r_{\mathrm{j}},s_{\mathrm{k}}) = \left[ \left( \frac{r_{\mathrm{j}}}{r_0} \right)^{-2\alpha_{\mathrm{in}}} + \left( \frac{r_{\mathrm{j}}}{r_0} \right)^{-2\alpha_{\mathrm{out}}} \right] ^{-1/2} \times N_{\mathrm{s}}(s_{\mathrm{k}}),
\end{equation}
where $N_{\mathrm{s}}(s_{\mathrm{k}})$ is a function depending solely on the grain size (similarly to Eq.\,2 of \citealp{Dullemond2008})
\begin{equation}
	N_{\mathrm{s}}(s_{\mathrm{k}}) = \left( \frac{s_{\mathrm{k}}}{s_{\mathrm{min}}} \right) ^{p} \times s_{\mathrm{k}} \times \Delta \mathrm{log}(s)_{\mathrm{k}},
\end{equation}
where $\Delta \mathrm{log}(s)_{\mathrm{k}}$ is the width of each bin in logarithmic space of $s$. Since the $n_{\mathrm{g}}$ grain sizes are logarithmically spaced, each $\Delta \mathrm{log}(s)_{\mathrm{k}}$ is the same, except the first and last ones which are half of that value. The mass density distribution is then given by $M_{\mathrm{dens}} (r_{\mathrm{j}}, s_{\mathrm{k}}) = N_{\mathrm{dens}} (r_{\mathrm{j}}, s_{\mathrm{k}}) \times 4/3 \pi s_{\mathrm{k}}^3 \rho$. Both $N_{\mathrm{dens}}(r_{\mathrm{j}},s_{\mathrm{k}})$ and $M_{\mathrm{dens}}(r_{\mathrm{j}},s_{\mathrm{k}})$ are then normalized so that the integral of $M_{\mathrm{dens}}(r_{\mathrm{j}},s_{\mathrm{k}}) \times V_{\mathrm{j}}$ (where $V_{\mathrm{j}}$ is the volume of the shell $\mathrm{j}$) over $s$ and $r$ equals the total input dust mass $M_{\mathrm{disk}}$.

The radially integrated optical depth at the wavelength of the peak of the stellar emission ($\tau_{\lambda\,\mathrm{peak}}$) is computed at the outer edge $r_{\mathrm{out}}$, assuming a spherical symmetry of the disk. This quantity is then divided by $f_{\psi} =  \mathrm{sin}(\psi)$ to obtain the final optical depth of the model, for a given $\psi$. This $1/f_{\psi}$ factor accounts for the fact that the code is 1D, i.e., that the disk is not a sphere but has a maximum vertical height $h$ from the midplane (hence can be seen as dust belts). In Fig.\,\ref{fig:ddit}, the shells are shown in fading colors, while the dust belts are represented without transparency. The $f_{\psi}$ factor is the ratio of the volume of a sphere of radius $R$ to which we subtract two spherical cones (up and down) of height $R \times (1 - \mathrm{sin}(\psi))$, divided by the volume of a sphere of the same radius. One spherical cone has a volume of $2\pi/3 R^3(1-\mathrm{sin}(\psi))$, which simplifies the volume ratio to $\mathrm{sin}(\psi)$. By doing so, we can keep $M_{\mathrm{disk}}$ constant and leave the output of the code unchanged while simply updating the optical depth. Once the model has been computed, a check is performed to see if $\tau_{\lambda\,\mathrm{peak}}$ is smaller than unity for the given dust mass and $\psi$.

When computing the SED, we consider that the contribution of scattered light is negligible at mid to far-IR wavelengths compared to the thermal emission. Furthermore, since debris disks are optically thin at all wavelengths, geometrical parameters such as the inclination have no impact on the final SED. Finally, as the observed SED is constructed from spatially unresolved observations, disk or aperture sizes do not have to be taken into consideration.

For each shell, the temperature of a dust grain is computed as a function of its size $s$ and distance $r$ to the star. Assuming radiative equilibrium, one can express $r$ as
\begin{equation}
r(s_{\mathrm{k}}, T_\mathrm{dust}) = \frac{R_{\star}}{2} \times \sqrt \frac{\int_{\lambda} F_{\star}(\lambda) Q_{\mathrm{abs}}(\lambda, s_{\mathrm{k}}) \mathrm{d}\lambda}{\int_{\lambda} \pi B_{\lambda}(\lambda, T_{\mathrm{dust}}) Q_{\mathrm{abs}}(\lambda, s_{\mathrm{k}}) \mathrm{d}\lambda},
\end{equation}
where $R_{\star}$ is the stellar radius, $F_{\star}$ the stellar surface flux, and $B_{\lambda}$ the Planck function at the dust temperature $T_{\mathrm{dust}}$. By inverting numerically the above equation, we obtain the radial dependency of the temperature. Afterwards, the flux $F_{\mathrm{th}}$ (the thermal radiation of a single dust grain) received by an observer at distance $d_{\star}$ is computed as
\begin{equation}\label{eqn:Fth}
F_{\mathrm{th}} (\lambda, r_{\mathrm{j}}, s_{\mathrm{k}}) = \left( \frac{4 \pi s_{\mathrm{k}}^2}{4 \pi d_{\star}^2} \right) \pi B_{\lambda}(\lambda, T_\mathrm{dust,j}) Q_{\mathrm{abs}}(\lambda, s_{\mathrm{k}}).
\end{equation}
For a given shell, the total thermal emission is obtained by multiplying $F_{\mathrm{th}} (\lambda, r_{\mathrm{j}}, s_{\mathrm{k}})$ by the normalised $N_{\mathrm{dens}} (r_{\mathrm{j}}, s_{\mathrm{k}})$. The final SED is the sum of the thermal emission from all $n_{\mathrm{r}}$ shells.

\subsection{Computing synthetic images}

\begin{figure}
\includegraphics[width=1.0\columnwidth]{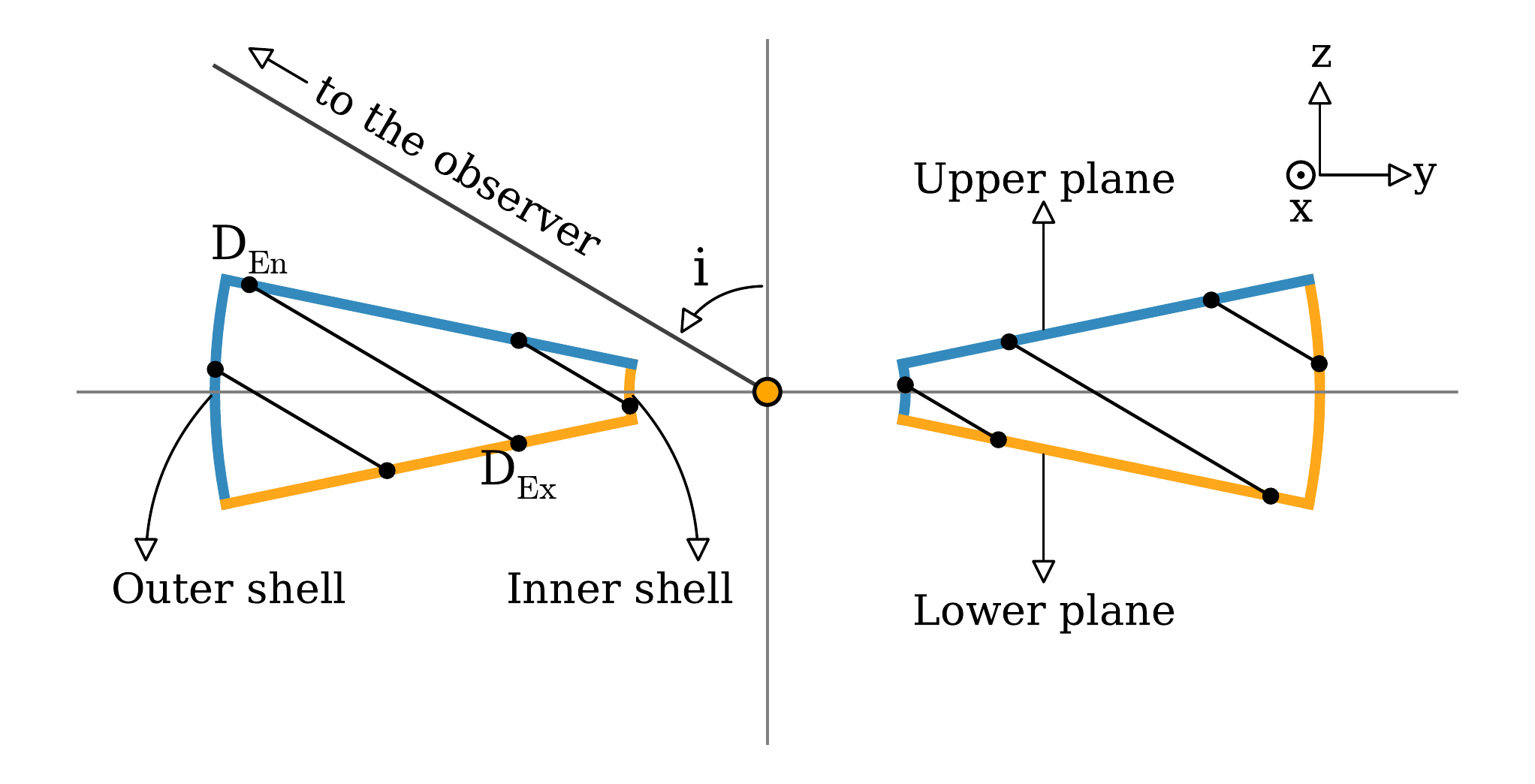}
\caption{Sketch of a disk (inclined by $i$) as seen from the side to illustrate the entry and exit points ($D_\mathrm{En}$ and $D_\mathrm{Ex}$) when computing images. The cut shows the $y$ and $z$ in the disk frame axis and is made at $x = 0$ (the $x$ axis pointing toward the reader).}
\label{fig:rays}
\end{figure}

At optical or near-IR wavelengths, stellar light scattered by dust grains can no longer be considered negligible and has to be accounted for. We follow a parallel approach as for the thermal emission and the stellar light scattered by a single dust grain is computed as follows
\begin{equation}\label{eqn:Fscat}
F_{\mathrm{scat}} (\lambda, r_{\mathrm{j}}, s_{\mathrm{k}}, \theta) = \frac{F_{\star} (\lambda)}{4\pi r_{\mathrm{j}}^2} \pi s_{\mathrm{k}}^2 \left( \frac{1}{4 \pi d_{\star}^2} \right) Q_{\mathrm{sca}}(\lambda, s_{\mathrm{k}}) S_{11} (\lambda, s_{\mathrm{k}}, \theta),
\end{equation}
where $Q_{\mathrm{sca}}$ is the scattering efficiencies calculated with the Mie theory and $S_{11}(\theta)$ is the scattering phase function.

In the model parametrization, we define two quantities for the image itself, the number of pixels $n_{\mathrm{p}}$ (images are squares) and the size $l_\mathrm{p}$ of one pixel in units of au. Then, for each of the $n_{\mathrm{p}}^2$ pixels in the image, we first apply adequate projection and rotation to account for the geometrical parameters $i$ and $\phi$ (inclination and position angle, respectively). For $\phi = 0^{\circ}$, the semi-major axis is along the south-north direction, and the front side of the disk is chosen to be on the east side. For increasing $\phi$ the disk rotates counter-clockwise, towards the east.

Since the disk has a non-zero opening angle $\psi$, it is not infinitely flat in the vertical direction. As in other codes (e.g. \textsc{GRaTeR}, \citealp{Augereau1999}), we assume a vertical Gaussian profile for the dust distribution, with a $\sigma$ width equal to $\psi$. This avoids having blocky images, especially for highly inclined disks. Figure\,\ref{fig:rays} shows a crude sketch of a cut in the $(y,z)$ reference plane of the disk, for $x = 0$. The disk has an inclination $i$ and for each pixel, we trace rays along the line of sight and search for intersections with the disk. The disk can be described by four main geometrical shapes: inner and outer shells, and upper and lower planes. Each of them can be either ``entry'' or ``exit'' points for the rays (blue and yellow lines in Fig.\,\ref{fig:rays}, respectively). One should note that only slightly different intersection conditions are to be considered for cases where $\pi/2 - i <  \psi$ (e.g., edge-on disk) and that the code can produce images for such disks. We can therefore obtain the coordinates of both the entry and exit points ($D_\mathrm{En}$ and $D_\mathrm{Ex}$, respectively). Since a pixel has a size $l_\mathrm{p}$, the column between $D_\mathrm{En}$ and $D_\mathrm{Ex}$ (modulus of $|l|$) has a volume of $l_{\mathrm{p}}^2 \times |l|$. Since this column can probe a significant range of distances $r$ to the star, we divide it into $n_{\mathrm{los}}$ boxes, along the line the sight. For each of these boxes, we evaluate the distance to the star $r$, the volume of the box $V_{\mathrm{box}}$, and the scattering angle (dot product between the line of sight and the position in the disk with respect to the star). We then interpolate $F_{\mathrm{th}}$, $F_{\mathrm{scat}}$, and $N_{\mathrm{dens}}$ at the distance $r$ of the center of the box, and interpolate the phase function at the scattering angle. We then sum the product $F_{\mathrm{th/scat}} \times N_{\mathrm{dens}} \times V_{\mathrm{box}}$ for each box, to obtain both the scattered light and thermal emission for this given pixel.

\subsubsection{Polarized images}

The code can also produce polarized light images on top of the total intensity images. Neglecting multiple scattering events, a reasonable assumption in optically thin disks, we compute $F_{\mathrm{pol}} (\lambda, r_{\mathrm{j}}, s_{\mathrm{k}}, \theta)$ similarly to Eq.\,\ref{eqn:Fscat} simply replacing $S_{11}$ by $S_{12}$. The second element of the M\"uller matrix $S_{12}$ is calculated using the Mie theory (while $S_{11}$ can be approximated by the Henyey-Greenstein function). We can therefore produce synthetic images along the Stokes vector $Q$, which can be used to model polarimetric observations.

\begin{figure}
\includegraphics[width=1.0\columnwidth]{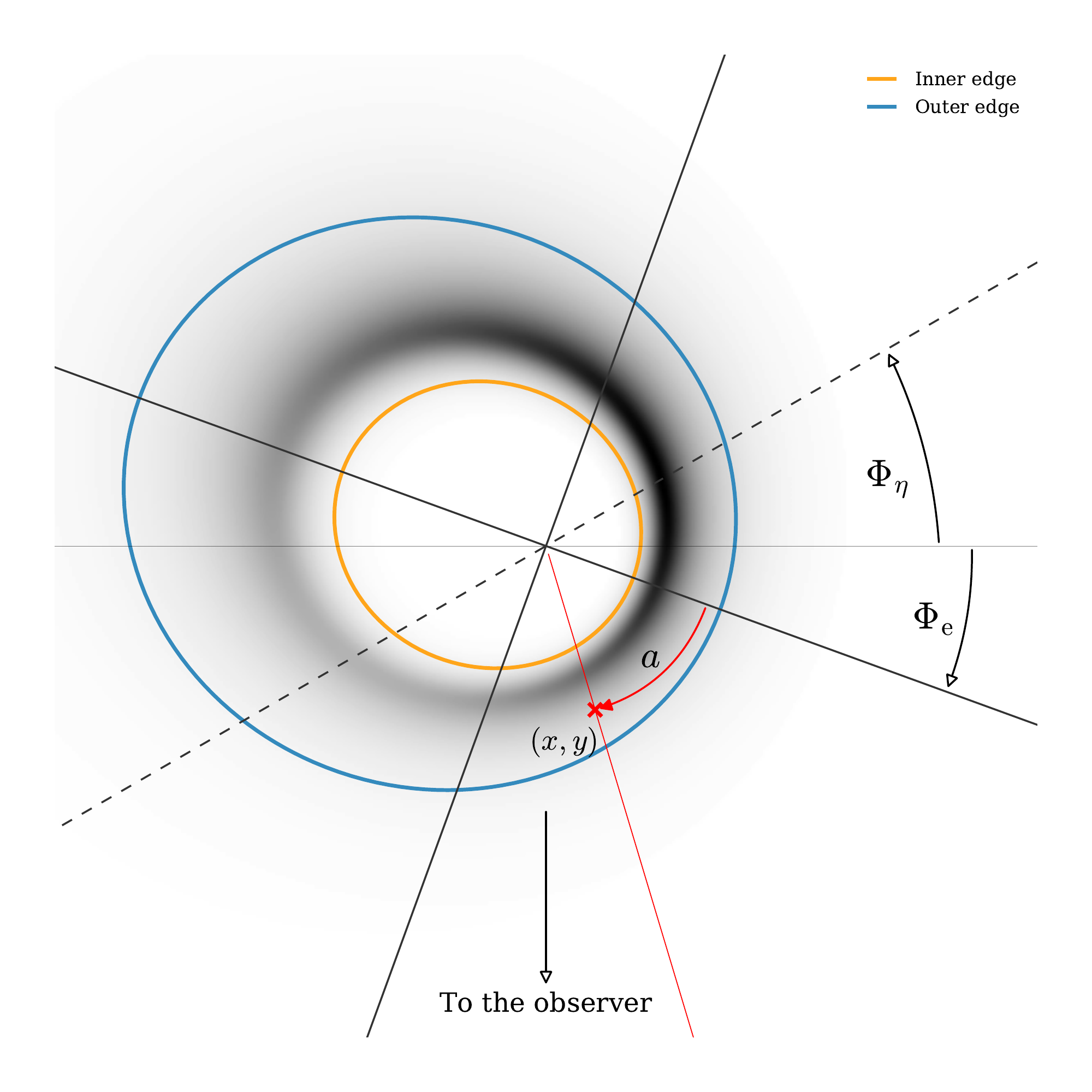}
\caption{Sketch of an eccentric disk. The color-coding represents the density distribution. The rotation angle is indicated by $\Phi_{\mathrm{e}}$, the reference angle for the azimuthal variation by $\Phi_{\eta}$, and $\eta = 0.3$ in this example.}
\label{fig:ecc}
\end{figure}

\subsubsection{Eccentric disks}

The code can produce images of eccentric disks. The inner and outer shells as described in Fig.\,\ref{fig:rays} are modified to inner and outer edges with an eccentricity $e$ parameter and a rotation angle $\Phi_{\mathrm{e}}$ (see Fig.\,\ref{fig:ecc} for a sketch of a face-on disk). For each pixel in the image, at position $(x,y)$ (in the reference frame rotated by $\Phi_{\mathrm{e}}$), we check if it is in between the inner and outer edges, which are defined as,
\begin{equation}\label{eqn:ecc}
\mathrm{Edge}_{\,\mathrm{inner,\,outer}} = \frac{r_{\mathrm{in,\,out}}}{1 + e \times \mathrm{cos}[a + \Phi_{\mathrm{e}}]},
\end{equation}
where $a = \mathrm{atan2}(y, x)$. For $\Phi_{\mathrm{e}} = \phi = 0^{\circ}$, the pericenter is located on the south side.

\subsubsection{Azimuthal density variations}

It is also possible to introduce azimuthal dust density variations in the synthetic images. The density distribution is parametrized with three parameters: a damping factor $\eta$ ($0 \leq \eta \leq 1$), a reference angle $\Phi_{\eta}$, and the azimuthal variation follows a Gaussian profile of $\sigma = w$ that peaks at $1$ for $\mathrm{atan2}(y, x) = \Phi_{\eta}$ and has an amplitude of $1 - \eta$. The attenuation factor is of the form
\begin{equation}
 N_{\mathrm{att}}(x, y) = \eta + (1 - \eta) \times \mathrm{exp}\frac{-(\mathrm{atan}(y,x) - \Phi_{\eta})^2}{2w^2}.
\end{equation}
For $\Phi_{\eta} = \phi = 0^{\circ}$, the density maximum is located on the south side. For a given azimuthal angle, $N_{\mathrm{dens}}$ is multiplied by $N_{\mathrm{att}}$ before computing the image.

\section{Results}

\begin{figure*}
\includegraphics[width=2\columnwidth]{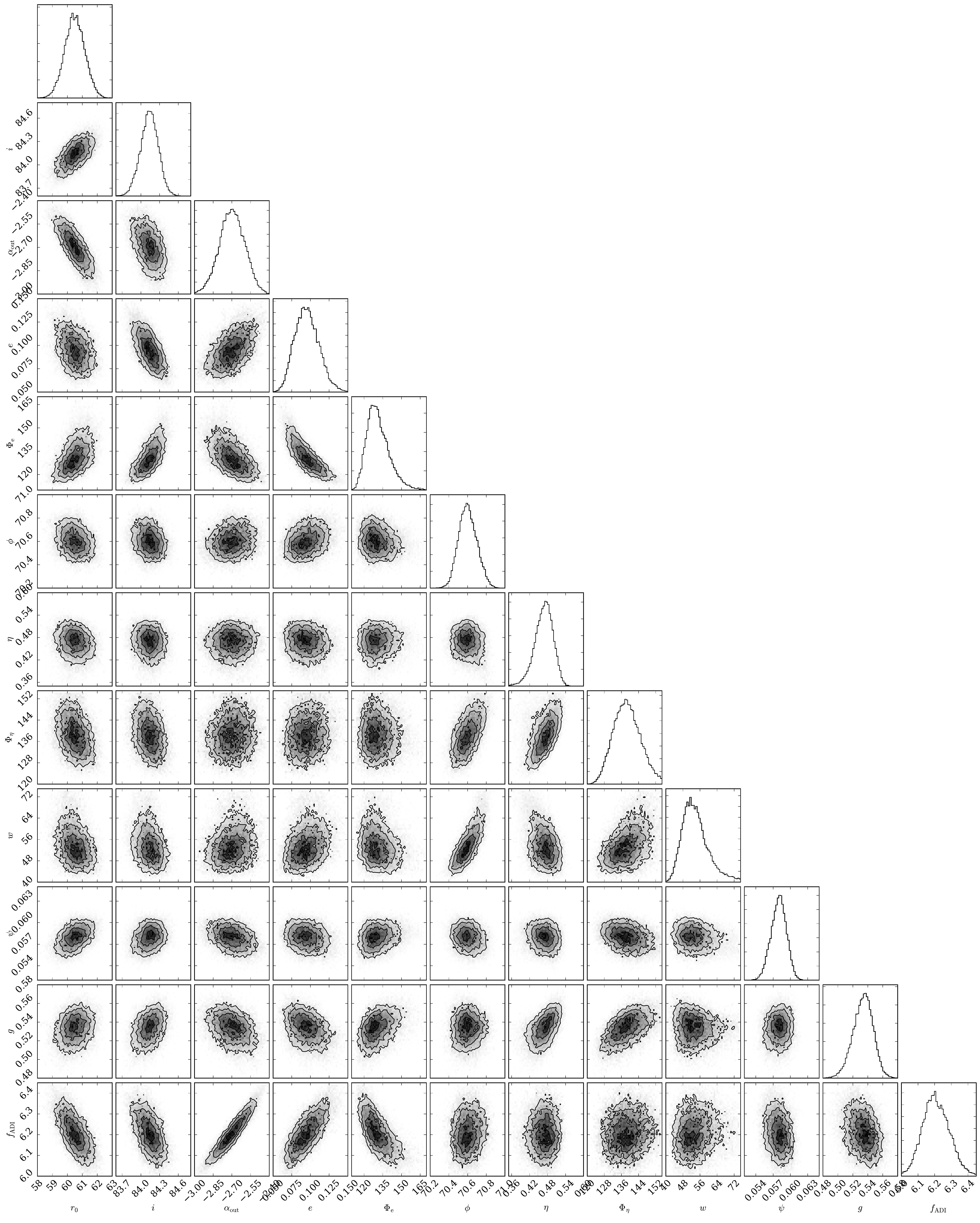}
\caption{Projected posterior probability distributions for the combined modeling of the ADI and DPI $H$-band observations.}
\label{fig:pdf_H}
\end{figure*}

\begin{figure}
\includegraphics[width=1.0\columnwidth]{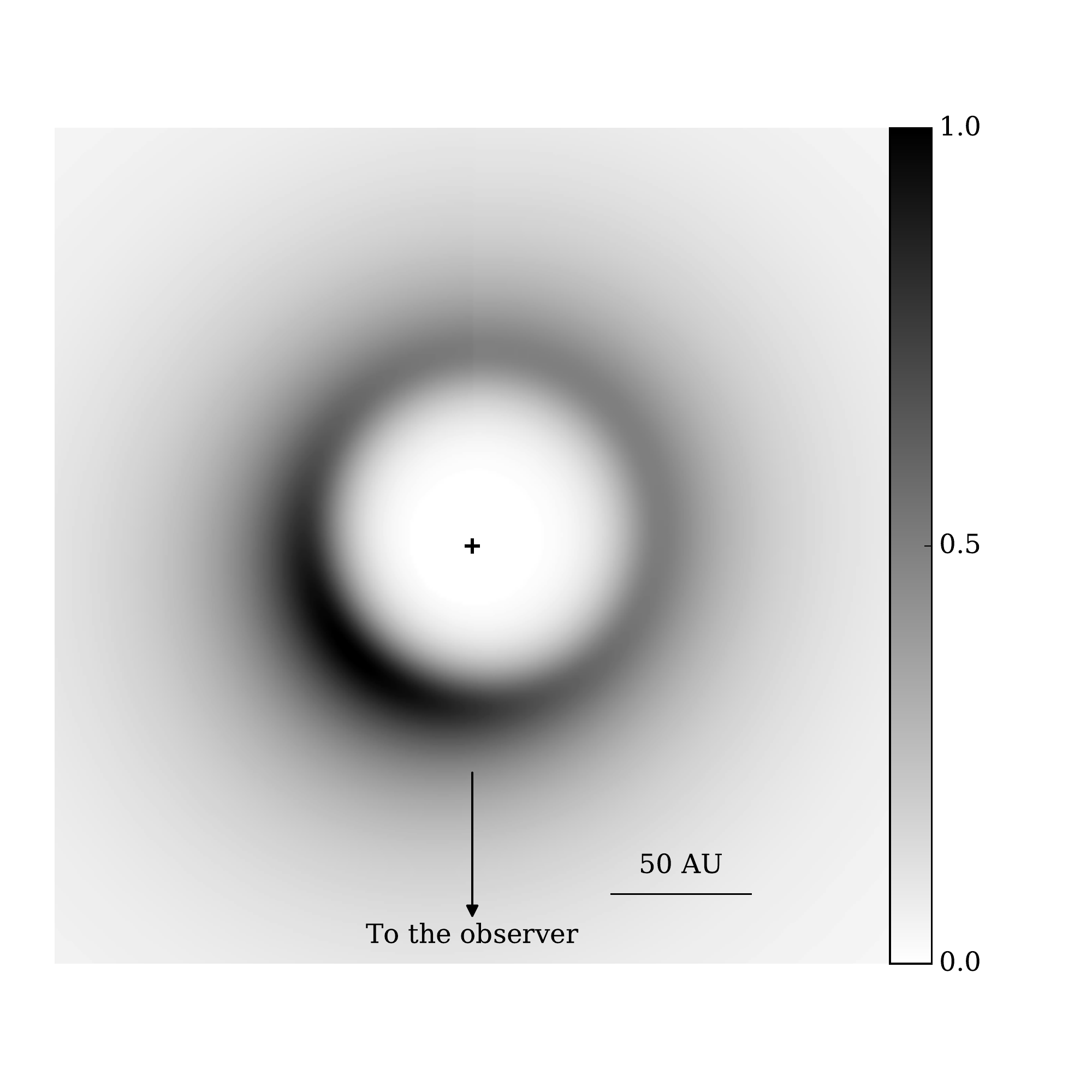}
\caption{Top view of the dust density distribution (normalized to its maximum) for the best-fit model to the SPHERE observations. The direction to the observer is indicated by the arrow and the spatial scale in the lower right corner.}
\label{fig:best_sketch}
\end{figure}

\section{Miscellaneous}

\begin{figure}
\includegraphics[width=1.0\columnwidth]{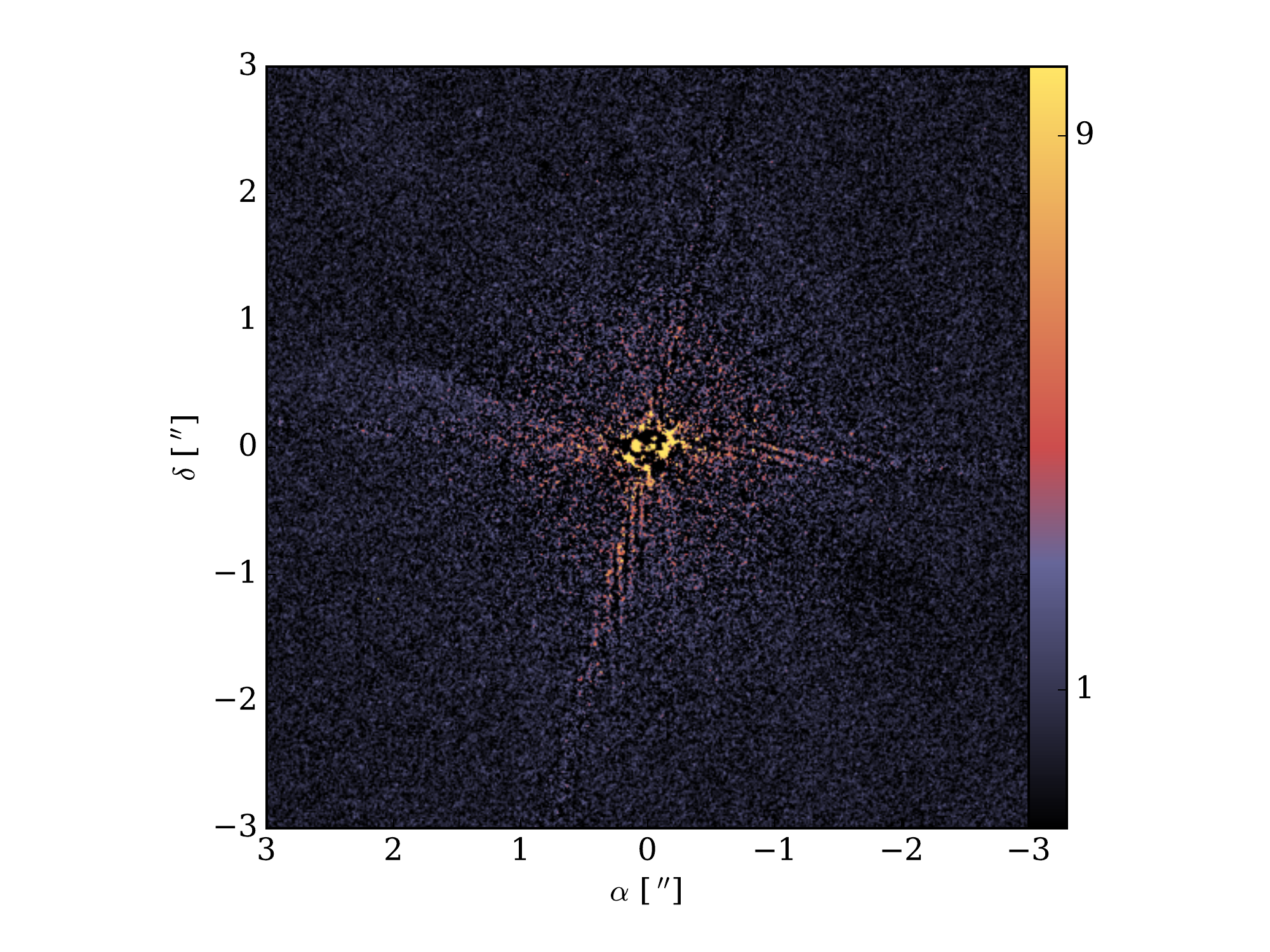}
\caption{$U_{\phi}$ component of the $H$-band DPI observations.}
\label{fig:uphi}
\end{figure}

\begin{figure}
\includegraphics[width=1.0\columnwidth]{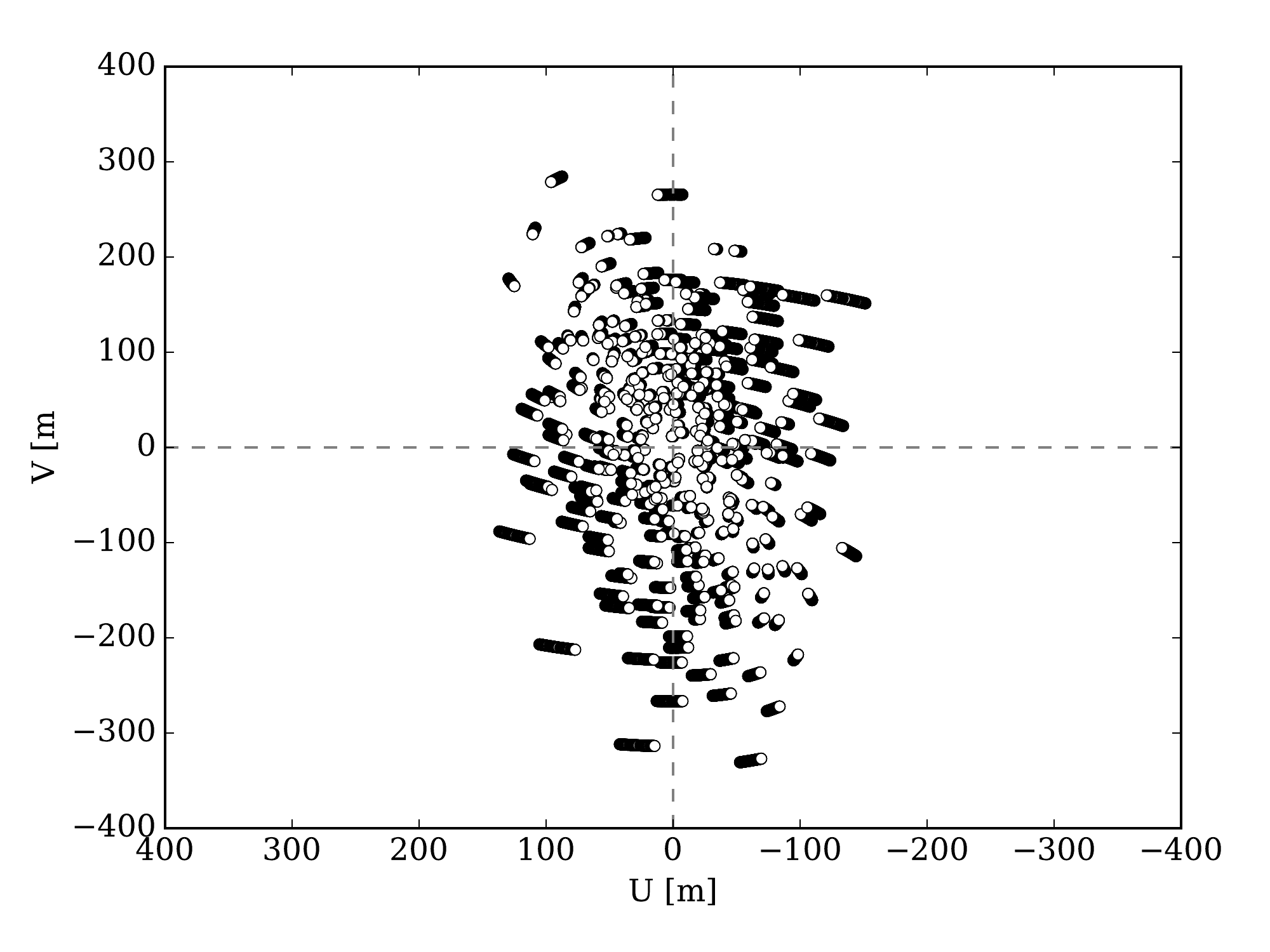}
\caption{Coverage of the (u,v) plane for the ALMA observations used in this paper.}
\label{fig:uvplane}
\end{figure}

\begin{figure}
\includegraphics[width=1.2\columnwidth]{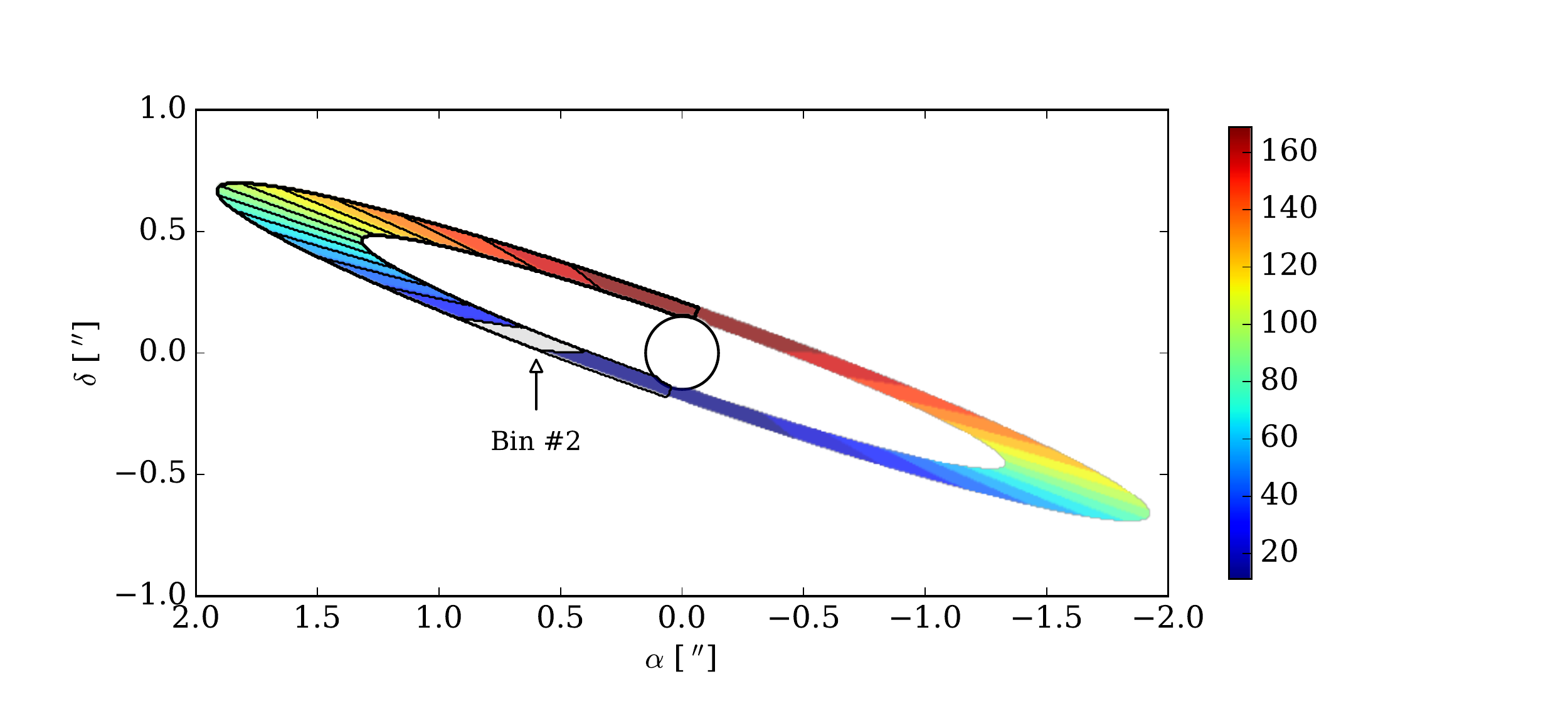}
\caption{Elliptical mask binned over different scattering angles to depict how the phase function is calculated. The second bin of the phase function would be the mean flux in the region highlighted in white. The color coding shows the scattering angle for a given position in the elliptical mask. The central circle shows the numerical mask to hide the central region.}
\label{fig:ell_pfunc}
\end{figure}

\begin{table*}
\centering
\caption{Notations used in this study.\label{tab:notations}}
\begin{tabular}{@{}lcr@{}}
\hline\hline
Notations & Units & Explanation \\
\hline
\hline
$L_{\star}$ & $[L_{\odot}]$ & Stellar luminosity \\
$M_{\star}$ & $[M_{\odot}]$ & Stellar mass \\
$R_{\star}$ & $[R_{\odot}]$ & Stellar radius \\
$T_{\star}$ & [K] & Stellar effective temperature \\
$d_{\star}$         & [pc] & Stellar distance \\
$i$ & [$^{\circ}$] & Inclination of the disk \\
$\phi$ & [$^{\circ}$] & Position angle of the disk (disk aligned North-South for $\phi = 0^{\circ}$, front side is towards the East) \\
$r_{0}$             & [au] & Semi latus rectum, the semi-major axis of the disk being $r_0 / (1-e^2)$ \\
$e$                 &      & Eccentricity of the disk \\
$\eta$              &      & Dust density damping factor \\
$\Phi_{\mathrm{e}}$ & [$^{\circ}$] & \specialcell{Reference angle to define the pericenter (along the semi-major\\axis, at the southern side for $\Phi_{\mathrm{e}} = \phi = 0^{\circ}$)} \\
$\Phi_{\eta}$ & [$^{\circ}$] & \specialcell{Reference angle for the peak density (along the semi-major\\axis, at the southern side for $\Phi_{\eta} = \phi = 0^{\circ}$)} \\
$w$ & [$^{\circ}$] & Width of the Gaussian profile for the azimuthal dust density variation\\
$g$ & & Asymmetry parameter for the HG phase function \\
$\alpha_{\mathrm{in}}$ &      & Slope for the inner dust density distribution \\
$\alpha_{\mathrm{out}}$ &     & Slope for the outer dust density distribution \\
$\psi$ & [rad] & Opening angle of the disk \\
$\beta$ &  & Radiation pressure to gravitational forces ratio \\
$s_{\mathrm{blow}}$ & [$\umu$m] & Dust grain size for which $\beta = 0.5$ \\
$s_{\mathrm{min}}$ & [$\umu$m] & Minimum grain size used in the model\\
$s_{\mathrm{max}}$ & [mm] & Maximum grain size \\
$p$ & - & Grain size distribution exponent \\
$f_{1300}$ & [mJy] & Flux at 1.3\,mm \\
$f_{\mathrm{ice}}$ & & Fraction of water ice mixed to the dust opacity\\
$f_{\mathrm{porosity}}$ & & Fraction of vacuum mixed to the dust opacity\\
M$_{\mathrm{dust}}$ & [M$_{\oplus}$] & Dust mass \\
$f_{\mathrm{ADI}}$ & [arbitrary] & Scaling factor for the ADI images \\
\hline
\end{tabular}
\end{table*}

\end{document}